\documentclass[hyper]{JHEP3}
\usepackage{amsmath,amssymb,amsfonts,epic,epsfig,psfrag,simplewick,time}
\DeclareMathAlphabet{\mathW}{OT1}{pnc}{m}{n}
\DeclareMathAlphabet{\mathsfsl}{OMS}{pplcm}{b}{n}
\author{V.~A.~Fateev$^{1,2}$ and A.~V.~Litvinov$^{1,3}$\\
$^1$~Landau Institute for Theoretical Physics, 142432 Chernogolovka, Russia.\\
$^2$~Laboratoire de Physique Th\'eorique et Astroparticules, UMR5207 CNRS-UM2, Universit\'e
Montpellier~II, Pl.~E.~Bataillon, 34095 Montpellier, France\\
$^3$~NHETC, Department of Physics and Astronomy, Rutgers University,\\ Piscataway, NJ 08855-0849, USA,\\
E-mail: \email{Vladimir.FATEEV@lpta.univ-montp2.fr},\email{litvinov@itp.ac.ru}}
\abstract{This is the second part of the paper \cite{Fateev:2007ab}. Here we show that three-point correlation function with one semi-degenerate field in Toda field theory as well as four-point correlation function with one completely degenerate and one semi-degenerate field can be represented by the finite dimensional integrals.}
\title{Correlation functions in conformal Toda field theory II}
\preprint{PTA/051\\RUNHETC-2008-16}
\dedicated{To the memory of Alyosha Zamolodchikov}
\begin{document}
\section{Introduction}
In this paper we continue our study of conformal field theories with extended symmetry which was started in the papers \cite{Fateev:2007ab,Fateev:2005gs} where $\mathfrak{sl}(n)$ Toda field theory was considered. This theory is given by the Lagrangian density
\begin{equation}\label{Lagrangian}
    \mathcal{L}=\frac{1}{8\pi}(\partial_a\varphi)^2+
    \mu\sum_{k=1}^{n-1} e^{b(e_k,\varphi)},
\end{equation}
here $\varphi$ is the two-dimensional $(n-1)$ component scalar field $\varphi=(\varphi_1\dots\varphi_{n-1})$, $b$ is the dimensionless coupling constant, $\mu$ is the scale parameter called the cosmological constant and $(e_k,\varphi)$ denotes the scalar product, where vectors $e_k$ are the simple roots of the Lie algebra $\mathfrak{sl}(n)$. Quantum field theory defined by the Lagrangian \eqref{Lagrangian} is known to be conformally invariant theory with additional symmetry ($W$-symmetry) generated by higher spin currents. Namely, there are $(n-1)$ holomorphic currents $\mathbf{W^{k}}(z)$ which form  closed $\mathW{W_n}$ algebra\footnote{There are different possible choices of the basis of holomorphic currents. In the paper \cite{Fateev:1987zh} this basis was defined from the Miura transformation, the advantage of this basis is that all commutation relations are bilinear. In some cases it is more convenient to have currents defined in such a way that they are primary with respect to stress-energy tensor. However this basis is not unique and some freedom in definition of higher currents is still remains. This freedom has to be fixed by additional requirements \cite{Bauer:1991ai}.}, which contains as subalgebra the Virasoro algebra with the central charge
\begin{equation}
    c=n-1+12Q^2=(n-1)(1+n(n+1)(b+b^{-1})^2).
\end{equation}
Basic objects of conformal Toda field theory are the exponential fields parameterized by a $(n-1)$ component vector parameter $\alpha$
\begin{equation}\label{field}
    V_{\alpha}=e^{(\alpha,\varphi)},
\end{equation}
which are the spinless primary fields. Important property of conformal Toda field theory is that two fields $V_{\alpha}$ and $V_{Q+\hat{s}(\alpha-Q)}$, where $\hat{s}$ is the element of Weyl group of the Lie algebra $\mathfrak{sl}(3)$ represent the same quantum field:
\begin{equation}\label{R_s}
    V_{Q+\hat{s}(\alpha-Q)}=R_{\hat{s}}(\alpha)V_{\alpha}\,,  
\end{equation}
where $R_{\hat{s}}(\alpha)$ is the reflection amplitude, which was found in \cite{Fateev:2001mj}
\begin{equation}\label{ReflAmp}
   \begin{gathered}
    R_{\hat{s}}(\alpha)=A(\alpha)/A(Q+\hat{s}(\alpha-Q)),\\
    A(\alpha)=(\pi\mu\gamma(b^2))^{\frac{(\alpha-Q,\rho)}{b}}
    \prod_{e>0}\Gamma(1-b(\alpha-Q,e))
    \Gamma(-b^{-1}(\alpha-Q,e)).
   \end{gathered}
\end{equation}
In eq \eqref{ReflAmp} the product goes over all positive roots. Other fields in the theory appear in the operator product of primary fields \eqref{field} with holomorphic currents $\mathbf{W^{k}}(z)$ and called descendant fields. Most important problem in conformal Toda field theory is to constract the whole set of multipoint correlation functions (of primary fields as well as of descendant fields). In order to solve this problem using the ideas of the operator algebra one has to find its structure constants of operator product expansion. In CFT with only Virasoro symmetry the properties of the operator algebra are significantly simplified and structure constants containing descendant fields can be obtained from the structure constants containing only primary fields \cite{Belavin:1984vu}. From this fact it follows immediately that in order to solve the theory completely one has to find structure constants for primary fields only or what is the same to find three-point correlation functions of primary fields. This statement in general is no longer true for the theories with additional symmetries ($\mathfrak{sl}(n)$ Toda field theory with $n>2$ for example), however it is still important problem to find three-point correlation functions of primary fields in these theories. Three-point function has universal coordinate behavior
\begin{equation}\label{Ctotal}
    \langle V_{\alpha_1}(z_1,\bar{z}_1)
    V_{\alpha_2}(z_2,\bar{z}_2)V_{\alpha_3}(z_3,\bar{z}_3)\rangle=\frac
    {C(\alpha_1,\alpha_2,\alpha_3)}
    {\vert z_{12}\vert^{2(\Delta_1+\Delta_2-\Delta_3)}
    \vert z_{13}\vert^{2(\Delta_1+\Delta_3-\Delta_2)}
    \vert z_{23}\vert^{2(\Delta_2+\Delta_3-\Delta_1)}}.
\end{equation}
Non-trivial component of \eqref{Ctotal} is constant $C(\alpha_1,\alpha_2,\alpha_3)$ which supposed to be very complicated function of parameters $\alpha_k$. It was shown in refs \cite{Fateev:2007ab,Fateev:2005gs} that function $C(\alpha_1,\alpha_2,\alpha_3)$ has clear analytical structure and can be expressed in terms of $\Upsilon$-function which was defined in \cite{Zamolodchikov:1995aa} by the integral representation
\begin{equation}
    \log\Upsilon(x)=\int_{0}^{\infty}\frac{dt}{t}
    \left[\left(\frac{b+b^{-1}}{2}-x\right)^2e^{-t}-\frac
    {\sinh^2\left(\frac{b+b^{-1}}{2}-x\right)\frac{t}{2}}
    {\sinh\frac{bt}{2}\sinh\frac{t}{2b}} 
    \right],
\end{equation} 
if one of the parameters $\alpha_k$ takes the special values (so called semi-degenerate fields). For example, if (up to Weyl transformation)
\begin{equation}
  \alpha_3=\varkappa\omega_{n-1}\qquad(\text{or}\quad\alpha_3=\varkappa\omega_{1}),
\end{equation}
where $\omega_{n-1}$ is the last fundamental weight of $\mathfrak{sl}(n)$ and $\varkappa$ is some numerical constant, then function $C(\alpha_1,\alpha_2,\varkappa\omega_{n-1})$ is given by the simple expression
\begin{multline}\label{C}
    C(\alpha_1,\alpha_2,\varkappa\omega_{n-1})=
    \left[\pi\mu\gamma(b^2)b^{2-2b^2}\right]^
    {\frac{(2Q-\sum\alpha_i,\rho)}{b}}\times\\\times
    \frac{\left(\Upsilon(b)\right)^{n-1}\Upsilon(\varkappa)
    \prod\limits_{e>0}\Upsilon\Bigl((Q-\alpha_1,e)\Bigr)
    \Upsilon\Bigl((Q-\alpha_2,e)\Bigr)}
    {\prod\limits_{ij}\Upsilon\Bigl(\frac{\varkappa}{n}+
    (\alpha_1-Q,h_i)+(\alpha_2-Q,h_j)\Bigr)}.
\end{multline}
The product in the numerator runs over all positive roots and in the denominator over  the weights $h_k$ of the first fundamental representation $\pi_1$ of the Lie algebra $\mathfrak{sl}(n)$ with the highest weight $\omega_{1}$ (first fundamental weight)
\begin{equation}
    h_{k}=\omega_{1}-e_{1}-\dots-e_{k-1}.
\end{equation}
The same is true for $\alpha_3=\varkappa\omega_1$, in this case expression for three-point correlation function  is given by \eqref{C} but with $h_k\rightarrow -h_{k}$. More general structure constants have rather complicated analytical structure and are not known at the moment in general form. In the paper \cite{Fateev:2007ab} we also considered three-point correlation function \eqref{Ctotal} in semiclassical limit $b\rightarrow0$ with
"light" parameters
\begin{equation}
    \alpha_k=b\eta_k,
\end{equation}
and in minisuperspace approximation with 
\begin{equation}
    \alpha_1=Q+ibP_1,\qquad\alpha_2=Q+ibP_2\quad\text{and}\quad\alpha_3=bs.
\end{equation}
In both cases we showed that three-point correlation function can be expressed in terms of finite-dimensional Barnes-like integral. It was shown that if one of the fields is semi-degenerate then in both cases these integrals can be performed. In this paper we describe more general structure constants which can be written in terms of finite dimensional Coulomb-like integrals. For simplicity we consider $\mathfrak{sl}(3)$ case. We find structure constant with arbitrary $\alpha_1$ and $\alpha_2$ and with $\alpha_3=\varkappa\omega_2-mb\omega_1$ corresponding to semi-degenerate field where $m$ is non-negative integer number\footnote{The representation of the $W$-algebra corresponding to the semi-degenerate field with parameters $\alpha_3=\varkappa\omega_2-mb\omega_1$ contains only one null-vector at the level $m+1$ contrary to the completely degenerate field (see below) which contains at least two independent null-vectors in the corresponding representation.} and show that this three-point correlation function can be written in terms of Coulomb integral of dimension $4m$.

Another interesting object is four-point correlation function with one completely degenerate field (the simplest completely degenerate field is $V_{-b\omega_1}$)
\begin{equation}\label{4point-definition}
   \langle V_{-b\omega_1}(z,\bar{z})V_{\alpha_1}(z_1,\bar{z}_1)
    V_{\alpha_2}(z_2,\bar{z}_2)V_{\alpha_3}(z_3,\bar{z}_3)\rangle.
\end{equation}
Contrary to the $\mathfrak{sl}(2)$ case this function does not satisfy Fuchsian differential equation of the order $n$ \cite{Fateev:2005gs,Bowcock:1993wq} and it seems that it cannot be a solution of Fuchsian differential equation of any finite order. This obstruction is a principal difference between Liouville and Toda field theories. The point is that $W$-conformal block is not fixed completely by $W$-invariance contrary to Liouville case. In the paper \cite{Fateev:2007ab} we considered Toda field theory in  semiclassical limit $b\rightarrow0$ with "heavy" parameters
\begin{equation}
   \alpha_k=\frac{\eta_k}{b}.
\end{equation}
In this case semiclassical limit of four-point correlation function \eqref{4point-definition} satisfies differential equation of the order $n$ but this equation contains accessory parameters which have to be determined from the condition that four-point correlation function is single-valued. This condition gives transcendent equations for these parameters. Numerical analysis which was done for $\mathfrak{sl}(3)$ case in the collaboration with Enrico Onofri and will be published elsewhere shows that solution to these equations is unique only in special domain of parameters $\eta_1$, $\eta_2$ and $\eta_3$ (see ref \cite{Fateev:2007ab}). In this paper we show that in $\mathfrak{sl}(3)$ case if one of the fields is semi-degenerate (for example $\alpha_3=\varkappa\omega_2-mb\omega_1$) 
then quantum four-point correlation function can be expressed in terms of $(4m+4)$-dimensional Coulomb integral. For $m=0$  this integral is a solution of differential equation of the third order which is related with higher hypergeometric equation and can be expressed in terms of hypergeometric function $_3F(x)_2$ \cite{Fateev:2007ab}.

The plan of the paper is as follows. In section \ref{SL3-Lagrangian} we define basic notations concerning $\mathfrak{sl}(3)$ Toda field theory. In section \ref{3xtochka} we consider three-point correlation function and derive integral representation in the case when one of the fields is semi-degenerate (see eq \eqref{C-m}). We also prove identities for structure constants with degenerate fields announced in \cite{Fateev:2007ab}. In section \ref{4xtochka} we consider four-point correlation function with one degenerate field and one semi-degenerate field and also derive integral representation for this function (eq \eqref{4point-final}). In section \ref{Conclusion} we make concluding remarks and in the appendices we collect useful integral identities used in this paper.
%%%%%%%%%%%%%%%%%%%%%%%%%%%%%%%%%%%%%%%%%%%%%%%%%%%%%%%%%%%%%%%%%%%%%%%%%%%%%%%%%%%%%%%%%%%%%%%%%%%%%%%%%%%
%%%%%%%%%%%%%%%%%%%%%%%%%%%%%%%%%%%%%%%%%%%%%%%%%%%%%%%%%%%%%%%%%%%%%%%%%%%%%%%%%%%%%%%%%%%%%%%%%%%%%%%%%%%
%%%%%%%%%%%%%%%%%%%%%%%%%%%%%%%%%%%%%%%%%%%%%%%%%%%%%%%%%%%%%%%%%%%%%%%%%%%%%%%%%%%%%%%%%%%%%%%%%%%%%%%%%%%
\section{$\mathfrak{sl}(3)$ Toda field theory}\label{SL3-Lagrangian}
In this section and further in this paper we will consider $\mathfrak{sl}(3)$ Toda field theory. This theory on a surface with metric $\hat{g}_{ab}$ is described by the action
\begin{equation}\label{SL3_Action}
  S_{T}=\int\left(
  \frac{1}{8\pi}\hat{g}^{ab}(\partial_a\varphi,\partial_b\varphi)+
   \frac{(Q,\varphi)}{4\pi}\hat{R}+
   \mu\sum_{k=1,2}e^{b(e_k,\varphi)}
  \right)\sqrt{\hat{g}}\;d^2x,
\end{equation}
here $\hat{R}$ is the scalar curvature of the background metric, $\varphi$ is two-component quantum field $\varphi=(\varphi_1,\varphi_2)$. Vectors $e_1$ and $e_2$ are the simple roots with the matrix of scalar products $K_{ij}=(e_i,e_j)$:
\begin{equation}
   K_{ij}=
   \begin{pmatrix}
    2&-1\\
    -1&2
   \end{pmatrix}.
\end{equation}
Theory defined by the action \eqref{SL3_Action} can be viewed as a generalization of Liouville field theory widely considered in the literature due to its connection with strings in non-critical dimension \cite{Polyakov:1981rd}. Liouville field theory which is non-rational conformal field theory is governed by the Virasoro algebra. $\mathfrak{sl}(3)$ Toda field theory is governed by more involved symmetry algebra. Namely, the chiral part of the symmetry algebra of the theory contains two currents of the spin two and three
\begin{equation}\label{currents}
    \mathbf{W^2}(z)=T(z)=
    \sum_{n=-\infty}^{\infty}\frac{L_n}{z^{n+2}}\qquad
    \text{and}\qquad
    \mathbf{W^3}(z)=W(z)=
    \sum_{n=-\infty}^{\infty}\frac{W_n}{z^{n+3}}.
\end{equation}
The Laurent componets $L_k$ and $W_k$ form closed  Zamolodchikov's $\mathW{W_3}$ algebra with the commutation relations  \cite{Zamolodchikov:1985wn,Fateev:1987vh}
\begin{subequations}\label{W3algebra}
\begin{equation}\label{LunderL}
    \left[L_n,L_m\right]=(n-m)L_{n+m}+\frac{c}{12}(n^3-n)
    \delta_{n,-m},
\end{equation}
\begin{equation}\label{WunderL}
    \left[L_n,W_m\right]=(2n-m)W_{n+m},
\end{equation}
\begin{multline}\label{WunderW}
    \left[W_n,W_m\right]=\frac{c}{3\cdot5!}(n^2-1)(n^2-4)n
    \delta_{n,-m}+\frac{16}{22+5c}(n-m)\Lambda_{n+m}+\\+
    (n-m)\left(\frac{1}{15}(n+m+2)(n+m+3)-\frac{1}{6}(n+2)(m+2)
    \right)
    L_{n+m},
\end{multline}
\end{subequations}
here
\begin{equation*}
    \Lambda_n=\sum_{k=-\infty}^{\infty}:L_kL_{n-k}:+\frac{1}{5}x_n
    L_n,
\end{equation*}
\begin{equation*}
    x_{2l}=(1+l)(1-l)\qquad x_{2l+1}=(2+l)(1-l).
\end{equation*}
$\mathW{W_3}$ algebra defined by commutation relations \eqref{W3algebra} contains as subalgebra Virasoro algebra with central charge
\begin{equation}\label{C_T}
  c=2+24\left(b+\frac{1}{b}\right)^2.
\end{equation}
Primary fields of the theory $V_{\alpha}=e^{(\alpha,\varphi)}$ are the highest weight fields of $W$-algebra
\begin{equation}
   L_0V_{\alpha}=\Delta(\alpha)V_{\alpha},\qquad W_0V_{\alpha}=w(\alpha)V_{\alpha},
   \qquad L_{n}V_{\alpha}=W_{n}V_{\alpha}=0\qquad\text{for}\qquad n>0,
\end{equation}
where
\begin{subequations}
\begin{equation}\label{delta}
  \Delta(\alpha)=\frac{(2Q-\alpha,\alpha)}{2}
\end{equation}
is the conformal dimension and
\begin{equation}\label{omega}
  w(\alpha)=i\sqrt{\frac{48}{22+5c}}\;
  (\alpha-Q,h_1)(\alpha-Q,h_2)(\alpha-Q,h_3)
\end{equation}
\end{subequations}
is the quantum number associated to the $W(z)$ current. Other generators of the algebra $L_{-n}$ and $W_{-n}$ with $n>0$ create new fields which called descendant fields.  The quantum numbers \eqref{delta} and \eqref{omega} possess the symmetry under the action of the Weyl group $\mathsfsl{W}$ of the Lie algebra $\mathfrak{sl}(3)$ \cite{Fateev:1987zh}. This group is generated by the elements $\boldsymbol{\sigma_1}$ and $\boldsymbol{\sigma_2}$ which are reflections in the hyperplanes orthogonal to the simple roots $e_1$ and $e_2$
\begin{equation}\label{Sigma_k}
   \begin{aligned}
       &\boldsymbol{\sigma_1}(\alpha)=\alpha-(\alpha-Q,e_1)e_1,\\
       &\boldsymbol{\sigma_2}(\alpha)=\alpha-(\alpha-Q,e_2)e_2.
   \end{aligned}
\end{equation}

Quantum field theory defined by action \eqref{SL3_Action} despite its wide symmetry algebra is still very complicated, but some information about correlation functions can be obtained from the "zero-mode integration method" developed in ref \cite{Goulian:1990qr}. To proceed we consider the geometry of a sphere and fix metric $\hat{g}_{ab}=\delta_{ab}$ everywhere except the north pole $z=\infty$ where the curvature will be collected. Correlation functions of exponential fields $V_{\alpha}$ are defined as follows
\begin{equation}\label{n-point-functional-integral}
  \langle V_{\alpha_1}(z_1)\dots V_{\alpha_n}(z_n)\rangle_T=
  \int\left[\mathcal{D}\varphi\right]e^{-S_T[\varphi]}e^{(\alpha_1,\varphi(z_1))}
   \dots e^{(\alpha_n,\varphi(z_n))},
\end{equation}
where action is written in a fixed metric
\begin{equation}
   S_T[\varphi]=\int
   \left(\frac{1}{8\pi}(\partial_a\varphi)^2+\mu e^{b(e_1,\varphi)}+\mu e^{b(e_2,\varphi)}+
   \frac{1}{4\pi}R(Q,\varphi)\right)
   d^2\xi.
\end{equation}
We define the components $\varphi_1$ and $\varphi_2$ of the field $\varphi$ in the basis of fundamental weights $\omega_1$ and $\omega_2$ of the Lie algebra $\mathfrak{sl}(3)$
\begin{equation}
  \varphi=\varphi_1\omega_1+\varphi_2\omega_2
\end{equation}
and change variables for the future convenience
\begin{equation}
   \varphi_1=\phi\sqrt{3}-\Phi,\qquad \varphi_2=2\Phi.
\end{equation}
Our goal is to make integration in correlation function \eqref{n-point-functional-integral} over zero-mode $\phi_0$ of the field $\phi$
\begin{equation}
  \phi=\phi_0+\tilde{\phi}.
\end{equation}
As a result of integration we obtain
\begin{multline}\label{n-point-after-GL}
  \langle V_{\alpha_1}(z_1)\dots V_{\alpha_n}(z_n)\rangle_T=
  \frac{\mu^s\Gamma(-s)}{b}\times\\
  \int e^{-S_L[\Phi]}e^{-S_0[\tilde{\phi}]}
  \left(\int e^{b\sqrt{3}\tilde{\phi}-b\Phi}\right)^s
  e^{(\alpha_1,e_2)\Phi(z_1)+(\alpha_1,\omega_1)\sqrt{3}\tilde{\phi}(z_1)}\hspace{-2.51pt}
  \dots e^{(\alpha_n,e_2)\Phi(z_n)+(\alpha_n,\omega_1)\sqrt{3}\tilde{\phi}(z_n)},
\end{multline}
where 
\begin{equation}
  s=\frac{(2Q-\sum\alpha_k,\omega_1)}{b}.
\end{equation}
In eq \eqref{n-point-after-GL}
\begin{equation}\label{Liouville-action}
     S_L[\Phi]=\int\left(\frac{1}{4\pi}(\partial_a\Phi)^2+
      \mu e^{2b\Phi}+\frac{1}{4\pi}(b+b^{-1})\Phi R\right)d^2\xi
\end{equation}
is Liouville action with central charge  
\begin{equation}\label{QL}
  c_L=1+6Q_L^2
\end{equation}
where $Q_L=b+b^{-1}$ and
\begin{equation}\label{Free-action}
     S_0[\tilde{\phi}]=\int\left(\frac{1}{4\pi}(\partial_a\tilde{\phi})^2+
     \frac{\sqrt{3}}{4\pi}(b+b^{-1})\tilde{\phi}R\right)d^2\xi
\end{equation}
is free action with central charge $c_{\text{\tiny{FREE}}}=1+18Q_L^2$. Equality \eqref{n-point-after-GL} has no clear meaning if parameter $s$ is general, but for integer values of $s$ it relates residue of correlation function in Toda field theory with multiple integral containing special correlation function in Liouville field theory
\begin{multline}\label{Toda-Liouville-GL-relation}
  \text{Res}\Bigl|_{s=m}\langle V_{\alpha_1}(z_1)\dots V_{\alpha_n}(z_n)\rangle_T=(-\pi\mu)^m
  \prod_{i<j}^n|z_i-z_j|^{-3(\alpha_i,\omega_1)(\alpha_j,\omega_1)}\times\\
  \hspace*{-2.9pt}\int\prod_{i,j}^{m,n}|t_i-z_j|^{-3b(\alpha_j,\omega_1)}\prod_{i<j}^m|t_i-t_j|^{-3b^2}
  \hspace*{-2pt}\langle V_{-\frac{b}{2}}(t_1)...V_{-\frac{b}{2}}(t_m)
  V_{\frac{(\alpha_1,e_2)}{2}}(z_1)...V_{\frac{(\alpha_n,e_2)}{2}}(z_n)\rangle_L
  d\mu_m(t),
\end{multline}
where
\begin{equation}\label{dmu}
  d\mu_m(t)=\frac{1}{\pi^mm!}\prod_{j=1}^md^2t_j.
\end{equation}
Deriving equation \eqref{Toda-Liouville-GL-relation} we calculated the part of correlation function coming from the free theory with the action \eqref{Free-action} using standart Wick rules
\begin{equation}
   \acontraction{}{\phi}{(x)}{\phi}
   \phi(x)\phi(y)=-\log|x-y|.
\end{equation}
The non-trivial part of correlation function in the r.h.s. of \eqref{Toda-Liouville-GL-relation} labeled as $\langle\dots\rangle_L$ is calculated in the theory with Liouville action \eqref{Liouville-action}\footnote{Here we use a little bit missleading notations for the Liouville exponential fields. Namely, we define exponential field as $V_{\alpha}(z)=e^{2\alpha\Phi(z)}$ with conformal dimension $\Delta=\alpha(Q_L-\alpha)$ with $Q_L$ given by \eqref{QL}.}. 

Integral relation \eqref{Toda-Liouville-GL-relation} will be widely used in the following calculations for the case of three-point function. In this case in the r.h.s. of \eqref{Toda-Liouville-GL-relation} we have Liouville correlation function with three arbitrary and $m$-degenerate fields. Such correlation function  was studied in ref \cite{Fateev:2007qn} where the explicit integral representation for this correlation function was derived:
\begin{multline}\label{m-point}
\langle V_{-\frac{b}{2}}(t_1)\dots
  V_{-\frac{b}{2}}(t_m)V_{\frac{(\alpha_1,e_2)}{2}}(0)V_{\frac{(\alpha_2,e_2)}{2}}(1)
  V_{\frac{(\alpha_3,e_2)}{2}}(\infty)\rangle_L=\\=
  \Omega_m\left(\frac{(\alpha_1,e_2)}{2},\frac{(\alpha_2,e_2)}{2},\frac{(\alpha_3,e_2)}{2}\right)
  \prod_{k=1}^m\vert t_k\vert^{b(\alpha_1,e_2)}\vert t_k-1\vert^{b(\alpha_2,e_2)}
  \prod_{i<j}|t_i-t_j|^{-b^2}\times\\\times
  \int\prod_{k=1}^m\vert s_k\vert^{2A}\vert s_k-1\vert^{2B}K_m^{\boldsymbol{\Delta}}(s_1,..,s_m|t_1,..,t_m)
  \prod_{i<j}|s_i-s_j|^{-4b^2}\,d\mu_m(s)
\end{multline}
where
\begin{equation}\label{A-B-Delta}
    \begin{gathered}
       A=\frac{b}{2}\left((\alpha_2+\alpha_3-\alpha_1,e_2)+(m-2)b\right)-1,\quad
       B=\frac{b}{2}\left((\alpha_1+\alpha_2-\alpha_2,e_2)+(m-2)b\right)-1,\\
       \boldsymbol{\Delta}=\frac{b}{2}\left((\alpha_1+\alpha_2+\alpha_3,e_2)+(m-4)b\right)-1,
    \end{gathered}
\end{equation}
and normalization factor $\Omega_m(\lambda_1,\lambda_2,\lambda_3)$ is given by
\begin{multline}\label{O_m}
   \Omega_m(\lambda_1,\lambda_2,\lambda_3)=
   (-\pi\mu)^m
   \Bigl[\pi\mu\gamma(b^2)b^{2-2b^2}\Bigr]^{\frac{(Q-\lambda-mb/2)}{b}}\times\\\times
   \frac{\Upsilon'(-mb)\prod_{k=1}^3\Upsilon(2\lambda_k)}
    {\Upsilon(\lambda-Q-\frac{mb}{2})\prod_{k=1}^3\Upsilon(\lambda-2\lambda_k+\frac{mb}{2})},
\end{multline}
where $\lambda=\sum\lambda_k$. Function $K_m^{\boldsymbol{\Delta}}(s_1,..,s_m|t_1,..,t_m)$ was defined in the paper  \cite{Fateev:2007qn} (see also definition of this function in the appendix \ref{Kernel}). Finally we obtain that
\begin{multline}
   \text{Res}\Bigl|_{(2Q-\alpha,\omega_1)=mb}
   \langle V_{\alpha_1}(0)V_{\alpha_2}(1)V_{\alpha_3}(\infty)\rangle_T=\\=
   (-\pi\mu)^{m}\,
   \Omega_m\left(\frac{(\alpha_1,e_2)}{2},\frac{(\alpha_2,e_2)}{2},\frac{(\alpha_3,e_2)}{2}\right)
   \mathfrak{I}_m\left(\genfrac{}{}{0pt}{1}{A\;\;B}{A'\:B'}\,\boldsymbol{\Delta}\right),
\end{multline}
were function $\mathfrak{I}_m\left(\genfrac{}{}{0pt}{1}{A\;\;B}{A'\:B'}\,\boldsymbol{\Delta}\right)$
defined by the integral \eqref{5-parametric-integral}, parameters $A$, $B$ and $\boldsymbol{\Delta}$ are given by eq \eqref{A-B-Delta} and parameters $A'$ and $B'$ are given by
\begin{equation}
   A'=-b(\alpha_1,e_1),\;\;B'=-b(\alpha_2,e_1).
\end{equation}

Before proceed let us say few words about the structure of poles of correlation functions in TFT.
Namely, performing integration over zero-mode of the field $\varphi$ one gets that any multipoint correlation function of primary fields $\langle V_{\alpha_1}(z_1)\dots V_{\alpha_N}(z_N)\rangle$
exhibits a simple pole in variable $\alpha=\alpha_1+\dots+\alpha_N$ for the values
\begin{equation}\label{Scr-cond}
  (\alpha-2Q,\omega_1)=-mb\quad\text{or}\quad(\alpha-2Q,\omega_2)=-nb.
\end{equation}
Non-negative integer numbers $m$ and $n$ are called "the numbers of screening charges". If both conditions are satisfied the residue in the double pole\footnote{When both screening conditions \eqref{Scr-cond} are satisfied the correlation function has a poles in variables $(\alpha-2Q,\omega_1)$ and $(\alpha-2Q,\omega_2)$.} can be expressed in terms of free-field correlation functions
\begin{multline}\label{GoulLieFormula}
  \text{Res}_{(\alpha-2Q,\omega_1)=-mb}\text{Res}_{(\alpha-2Q,\omega_2)=-nb}
  \langle V_{\alpha_1}(z_1)\dots V_{\alpha_N}(z_N)\rangle_T=\\=
  \frac{(-\mu)^{m+n}}{m!\,n!}
  \langle V_{\alpha_1}(z_1)\dots V_{\alpha_N}(z_N)\mathcal{Q}_1^{m}\mathcal{Q}_{2}^{n}\rangle_0.
\end{multline}
In eq \eqref{GoulLieFormula} we have introduced notations for the so called screening charges
\begin{equation}\label{screening}
  \mathcal{Q}_{k}=\int e^{b(e_k,\varphi_k)}d^2\xi,\;\;k=1,2.
\end{equation}
In principle, due to symmetry of the theory under the change $b\rightarrow\frac{1}{b}$ correlation function $\langle V_{\alpha_1}(z_1)\dots V_{\alpha_N}(z_N)\rangle$ has poles in more general points
\begin{equation}\label{general-poles}
  (\alpha-2Q,\omega_1)=-mb-kb^{-1},\qquad (\alpha-2Q,\omega_2)=-nb-lb^{-1}.
\end{equation}
where non-negative integer numbers $k$ and $l$ represent the numbers of dual screening charges 
\begin{equation}\label{dual-screening}
  \tilde{\mathcal{Q}}_{k}=\int e^{b^{-1}(e_k,\varphi_k)}d^2\xi,\;\;k=1,2.
\end{equation}
This symmetry is a deep property of the theory originating from the fact that central charge \eqref{C_T} as well as quantum numbers of primary field \eqref{delta} and \eqref{omega} posses it. 
In fact the theory is symmetric under the change $b\rightarrow\frac{1}{b}$ and cosmological constant $\mu$ being replaced with 
\begin{equation}
   \tilde{\mu}=\frac{1}{\pi\gamma(1/b^2)}\left(
   \pi\mu\gamma(b^2)\right)^{1/b^2}.
\end{equation}
This symmetry can be verified in all cases when the exact answer is known.  Poles \eqref{general-poles} are not predicted from the classical Lagrangian description of the theory. Residue in the poles with $k$ and $l$ being both non-negative integers can be expressed in terms of more general Coulomb integrals which are more involved and usually defined by the contour integrals (see \cite{Dotsenko:1984nm,Dotsenko:1984ad}). Below to simplify the equations we will consider only the poles of the type \eqref{Scr-cond}.
%%%%%%%%%%%%%%%%%%%%%%%%%%%%%%%%%%%%%%%%%%%%%%%%%%%%%%%%%%%%%%%%%%%%%%%%%%%%%%%%%%%%%%%%%%%%%%%%%%%%%%%%%
%%%%%%%%%%%%%%%%%%%%%%%%%%%%%%%%%%%%%%%%%%%%%%%%%%%%%%%%%%%%%%%%%%%%%%%%%%%%%%%%%%%%%%%%%%%%%%%%%%%%%%%%%
%%%%%%%%%%%%%%%%%%%%%%%%%%%%%%%%%%%%%%%%%%%%%%%%%%%%%%%%%%%%%%%%%%%%%%%%%%%%%%%%%%%%%%%%%%%%%%%%%%%%%%%%%
\section{Three-point correlation function}\label{3xtochka}
Now let us consider first non-trivial case -- three-point correlation function \eqref{Ctotal}
\begin{equation}\label{3point}
   \langle V_{\alpha_1}(0)V_{\alpha_2}(1)V_{\alpha_3}(\infty)\rangle_T=C(\alpha_1,\alpha_2,\alpha_3).
\end{equation}
This function has simple poles when one of the conditions \eqref{general-poles} is satisfied.
It is natural to suppose, that these poles are the only simple poles of this function modulo Weyl transformation. For the future convenience we define a new function
\begin{multline}\label{C-improved}
   \mathfrak{C}(\alpha_1,\alpha_2,\alpha_3)=\\=
    \left[\pi\mu\gamma(b^2)b^{2-2b^2}\right]^{\frac{(\sum\alpha_i-2Q,\rho)}{b}}
    \frac{C(\alpha_1,\alpha_2,\alpha_3)}
    {\prod_{e>0}\Upsilon\bigl((Q-\alpha_1,e)\bigr)\Upsilon\bigl((Q-\alpha_2,e)\bigr)
    \Upsilon\bigl((Q-\alpha_3,e)\bigr)}.
\end{multline}
We note that function 
\begin{equation}
    Y(\alpha)=
    \left[\pi\mu\gamma(b^2)b^{2-2b^2}\right]^{\frac{(-\alpha,\rho)}{b}}
    \prod_{e>0}\Upsilon\bigl((Q-\alpha,e)
\end{equation}
satisfies reflection relations \eqref{R_s}
\begin{equation}
     Y(Q+\hat{s}(\alpha-Q))=R_{\hat{s}}(\alpha)Y(\alpha),
\end{equation}
where $R_{\hat{s}}(\alpha)$ is given by \eqref{ReflAmp} for $\mathfrak{sl}(3)$.

Function $\mathfrak{C}(\alpha_1,\alpha_2,\alpha_3)$ is symmetric and Weyl invariant function of variables $\alpha_1$, $\alpha_2$ and $\alpha_3$ which does not depend on cosmological constant $\mu$. It satisfies non-trivial functional relations (see appendix \ref{App-Integrals} for proof)
\begin{subequations}\label{Transformation2}
\begin{equation}
  \mathfrak{C}(\alpha_1,\alpha_2,\alpha_3)=
  \mathfrak{C}(\tilde{\alpha_1},\tilde{\alpha_2},\tilde{\alpha_3}),
\end{equation}
where
\begin{equation}\label{Transformation2.1}
    \tilde{\alpha_1}=\alpha_1-\varsigma_{ijk}h_i,\qquad
    \tilde{\alpha_2}=\alpha_2-\varsigma_{ijk}h_j,\qquad
    \tilde{\alpha_3}=\alpha_3-\varsigma_{ijk}h_k
\end{equation}
with 
\begin{equation}\label{sigma_{ijk}}
 \varsigma_{ijk}=(\alpha_1-Q,h_i)+(\alpha_2-Q,h_j)+(\alpha_3-Q,h_k),
\end{equation}
\end{subequations}
where $h_i$ are the weights of the first fundamental representation of Lie algebra $\mathfrak{sl}(3)$.
These relations allow us to simplify three-point correlation function in many cases. In particular, starting from the function 
\begin{equation}
  \mathfrak{C}(\alpha_1,\alpha_2,\varkappa\omega_2-mb\omega_1+\varepsilon\omega_1),
\end{equation}
where $\varepsilon$ is some infinitesimal number  and applying transformation \eqref{Transformation2} with $i=1$, $j=1$ and $k=2$ we obtain that
\begin{equation}
  \mathfrak{C}(\alpha_1,\alpha_2,\varkappa\omega_2-mb\omega_1+\varepsilon\omega_1)=
  \mathfrak{C}(\tilde{\alpha}_1,\tilde{\alpha}_2,\tilde{\alpha_3}),
\end{equation}
where
\begin{equation}\label{tilde(alpha_k)-1}
 \begin{aligned}
   &\tilde{\alpha}_1=\alpha_1-\left((\alpha_1-Q,h_1)+(\alpha_2-Q,h_1)+
    \frac{(\varkappa+mb-\varepsilon)}{3}\right)h_1,\\
   &\tilde{\alpha}_2=\alpha_2-\left((\alpha_1-Q,h_1)+(\alpha_2-Q,h_1)+
    \frac{(\varkappa+mb-\varepsilon)}{3}\right)h_1,\\
   &\tilde{\alpha}_3=\varkappa\omega_2-mb\omega_1+\varepsilon\omega_1
    -\left((\alpha_1-Q,h_1)+(\alpha_2-Q,h_1)+\frac{(\varkappa+mb-\varepsilon)}{3}\right)h_2.
 \end{aligned}
\end{equation}
One can easily check that in the limit $\varepsilon\rightarrow0$
\begin{equation}
   (\tilde{\alpha_1}+\tilde{\alpha_2}+\tilde{\alpha_3}-2Q,\omega_1)=-mb,
\end{equation}
so we meet the situation which is described by eq \eqref{Toda-Liouville-GL-relation}, i.e. function $\mathfrak{C}(\tilde{\alpha}_1,\tilde{\alpha}_2,\tilde{\alpha_3})$ with $\tilde{\alpha}_k$ given by \eqref{tilde(alpha_k)-1} has a pole with residue expressed in terms of integral with Liouville correlation function containing $m$ degenerate fields. 
From the other side function $\mathfrak{C}(\alpha_1,\alpha_2,\varkappa\omega_2-mb\omega_1+\varepsilon\omega_1)$ due to the factor $\Upsilon^{-1}((Q-\alpha_3,e_1))$ also has a pole at $\varepsilon\rightarrow0$
\begin{multline}
  \mathfrak{C}(\alpha_1,\alpha_2,\varkappa\omega_2-mb\omega_1+\varepsilon\omega_1)=
   \left[\pi\mu\gamma(b^2)b^{2-2b^2}\right]^
   {\frac{(\alpha_1+\alpha_2+\varkappa\omega_2-mb\omega_1-2Q,\rho)}{b}}\times\\\times
  \frac{1}{\varepsilon}\,
  \frac{1}{\Upsilon'(-mb)}
  \frac{C(\alpha_1,\alpha_2,\varkappa\omega_2-mb\omega_1)}
  {\Upsilon(\varkappa)\Upsilon((m+2)b+2b^{-1}-\varkappa)\prod_{e>0}
   \Upsilon\bigl((Q-\alpha_1,e)\bigr)\Upsilon\bigl((Q-\alpha_2,e)\bigr)}+O(1).
\end{multline}
Applying now eqs \eqref{Toda-Liouville-GL-relation}-\eqref{m-point} we obtain that correlation function with semi-degenerate field has a finite limit and can be expressed as
\begin{equation}\label{C-m}
  C(\alpha_1,\alpha_2,\varkappa\omega_2-mb\omega_1)=\Xi_m(\alpha_1,\alpha_2|\varkappa)\,\,
  \mathfrak{I}_m\left(\genfrac{}{}{0pt}{1}{A\;\;B}{A'\:B'}\,\boldsymbol{\Delta}\right),
\end{equation}
where integral $\mathfrak{I}_m\left(\genfrac{}{}{0pt}{1}{A\;\;B}{A'\:B'}\,\boldsymbol{\Delta}\right)$ is given by the eq \eqref{5-parametric-integral} with 
\begin{equation}
   A=\varrho_{32}^m,\;\;
   B=\varrho_{23}^m,\;\;
   A'=\varrho_{21}^m,\;\;
   B'=\varrho_{12}^m,\;\;
   \boldsymbol{\Delta}=1+\varrho_{22}^m,
\end{equation}
here parameters $\varrho_{ij}^m$ are given by
\begin{equation}\label{varrho_{ij}^m}
   \varrho_{ij}^m=-1-b^2+b\left(\frac{\varkappa+mb}{3}+(\alpha_1-Q,h_i)+(\alpha_2-Q,h_j)\right),
\end{equation}
and numerical factor $\Xi_m(\alpha_1,\alpha_2|\varkappa)$ by
\begin{multline}\label{Xi-m}
   \Xi_m(\alpha_1,\alpha_2|\varkappa)=(\pi\mu)^{2m}\left[\pi\mu\gamma(b^2)b^{2-2b^2}\right]^
  {\frac{(2Q-\alpha_1-\alpha_2-\varkappa\omega_2-mb\omega_1,\rho)}{b}}\times\\\times
  \frac{\Upsilon'(-mb)^2\Upsilon(\varkappa)\prod_{e>0}
  \Upsilon\bigl((Q-\alpha_1,e)\bigr)\Upsilon\bigl((Q-\alpha_2,e)\bigr)}
  {\prod_{ij}\Upsilon\Bigl(\frac{\varkappa+mb}{3}+(\alpha_1-Q,h_i)+(\alpha_2-Q,h_j)-mb\delta_{ij}\Bigr)}.
\end{multline}
Equation \eqref{C-m} is one of the main results of this paper. Using integral identities proved in the appendix \ref{Kernel} it can be shown that function $\mathfrak{I}_m\left(\genfrac{}{}{0pt}{1}{A\;\;B}{A'\:B'}\,\boldsymbol{\Delta}\right)$ can be reduced to the Coulomb integral of dimension $4m$\footnote{For $m=1$ this integral was explicitly calculated in \cite{Fateev:2007ab}}.

One of the important sets of fields in TFT is formed by completely degenerate fields \cite{Fateev:1987zh}. Completely degenerate fields  $V_{\alpha}$ in TFT are parameterized by two highest weights $\Omega_1$ and $\Omega_2$ of the finite dimensional representations of the Lie algebra $\mathfrak{sl}(3)$ and correspond to the value of the parameter $\alpha$ (up to Weyl transformation)
\begin{equation}\label{Completely_degenerate}
  \alpha=-b\Omega_1-\frac{1}{b}\Omega_2.
\end{equation}
These fields form closed operator algebra and posses an important property that in their operator product expansion with general primary field $V_{\alpha}$ appear only a finite number of primary fields $V_{\alpha'}$ with their descendant fields
\begin{equation}\label{Fusion_completely_degenerate}
   V_{-b\Omega_1-b^{-1}\Omega_2}V_{\alpha}=
   \sum_{s,p}C_{-b\Omega_1-b^{-1}\Omega_2,\alpha}^
   {\alpha'_{sp}}
   \left[V_{\alpha'_{sp}}\right],
\end{equation}
here by square brackets we denote the contribution of the descendant fields and introduce the parameter $\alpha'_{sp}$ as
\begin{equation}\label{Fusion_completely_degenerate-def}
  \alpha'_{sp}=\alpha-bh_s^{\Omega_1}-b^{-1}h_p^{\Omega_2}.
\end{equation}
In Eq \eqref{Fusion_completely_degenerate-def} $h_s^{\Omega}$ are the weights of the representation $\Omega$ and $C_{-b\Omega_1-b^{-1}\Omega_2,\alpha}^{\alpha'_{sp}}$ denotes the structure constant of the operator algebra. For simplicity we consider the case $\Omega_2=0$. Then we define structure constants for the field $V_{-b\Omega_1}$ with $\Omega_1=mb\omega_1+nb\omega_2$ with arbitrary field $V_{\alpha}$
\begin{equation}\label{Structure_Constant_light}
   \mathbb{C}_{m,\,n}^{k,\,l}(\alpha)=
   C_{-mb\omega_1-nb\omega_2,\alpha}^{\alpha-bh_{mn}^{kl}}.
\end{equation}
where
\begin{equation}\label{h-mn-kl}
  h_{mn}^{kl}=m\omega_1+n\omega_2-ke_1-le_2,
\end{equation}
and numbers $(k,l)$ lay inside a domain
\begin{equation}\label{Cond-nonzeros}
   k\geq0;\;\;l\geq0;\;\;\;\;\;\;m+n\geq k;\;\;m+n\geq l;\;\;\;\;\;\;
   n+k\geq l;\;\;m+l\geq k.
\end{equation}
This structure constant is given by the Coulomb integral
\begin{equation}
     \mathbb{C}_{m,\,n}^{k,\,l}(\alpha)=(-\pi\mu)^{k+l}
     \int\,
     \mathcal{I}_{kl}
     \left(\genfrac{}{}{0pt}{1}{-b(\alpha,e_1),\;\;-mg}{\hspace{-2pt}-b(\alpha,e_2),\;\;-ng}
     \Bigl|\genfrac{}{}{0pt}{1}{t}{s}\right)\,d\mu_k(t)\,d\mu_l(s),
\end{equation}
here and later we will use the following notation for the integrand
\begin{equation}
   \mathcal{I}_{kl}\left(\genfrac{}{}{0pt}{1}{A_1,\;\;B_1}{A_2,\;\;B_2}
   \Bigl|\genfrac{}{}{0pt}{1}{t}{s}\right)=
   \prod_{i=1}^k|t_i|^{2A_1}|t_i-1|^{2B_1}\mathcal{D}_k^{2g}(t)
   \prod_{j=1}^l|s_j|^{2A_2}|s_j-1|^{2B_2}\mathcal{D}_l^{2g}(s)
   \prod_{ij}|t_i-s_j|^{-2g},
\end{equation}
where $g=-b^2$, $d\mu_k(t)$ defined by eq \eqref{dmu} and
\begin{equation}
   \mathcal{D}_k(t)=\prod_{i<j=1}^k|t_i-t_j|^2.
\end{equation} 
Using relation \eqref{Transformation2} one can show that function 
\begin{multline}\label{StrConst-krasivaya}
    \Sigma_{m,\,n}^{k,\,l}(\alpha)=
    \frac{\left[\pi\mu\gamma(b^2)b^{2-2b^2}\right]^{-k-l}}
    {\Upsilon'(-mb)\Upsilon'(-nb)\Upsilon'(2b^{-1}+(m+n+2)b)}\times\\\times
    \frac{1}{\prod_{e>0}\Upsilon((Q-\alpha,e))\Upsilon((\alpha-bh_{mn}^{kl}-Q,e))}
    \,\mathbb{C}_{m,\,n}^{k,\,l}(\alpha)
\end{multline}
satisfies relations
\begin{equation}\label{Sigma-relations}
    \Sigma_{m,\,n}^{k,\,l}(\alpha)=\Sigma_{n,\,m}^{l,\,k}(\alpha)=
    \Sigma_{k,\,m+n-k}^{m,\,m+l-k}(\alpha-(m-k)bh_1)
    =\Sigma_{m+l-k,\,n+k-l}^{l,\,k}(\alpha-(l-k)bh_2).
\end{equation}
Using these relations one can always put index $N$ in \eqref{StrConst-krasivaya}  instead of index $m$, where $N$ is the minimum of numbers $m$, $n$, $k$, $l$, $(m+n-k)$, $(m+n-l)$, $(m+l-k)$ and $(n+k-l)$. As follows from the results of this section this structure constant can be always represented as an integral of dimension $4N$. We note that multiplicity of the weight $h_{mn}^{kl}$ equals to $N+1$. It means that the number of integration in correlation function depends on multiplicities. 
Relations \eqref{Sigma-relations} was discovered in \cite{Fateev:2007ab} in the light semiclassical level, where they were rather non-trivial.
%%%%%%%%%%%%%%%%%%%%%%%%%%%%%%%%%%%%%%%%%%%%%%%%%%%%%%%%%%%%%%%%%%%%%%%%%%%%%%%%%%%%%%%%%%%%%%%%%%%%%%%%%%%
%%%%%%%%%%%%%%%%%%%%%%%%%%%%%%%%%%%%%%%%%%%%%%%%%%%%%%%%%%%%%%%%%%%%%%%%%%%%%%%%%%%%%%%%%%%%%%%%%%%%%%%%%%%
%%%%%%%%%%%%%%%%%%%%%%%%%%%%%%%%%%%%%%%%%%%%%%%%%%%%%%%%%%%%%%%%%%%%%%%%%%%%%%%%%%%%%%%%%%%%%%%%%%%%%%%%%%%
\section{Four-point correlation function}\label{4xtochka}
In this section we consider four-point correlation function with one degenerate field. The importance of this object manifested itself already in the Liouville field theory where it was very effective tool for the solution of the associativity condition of the operator algebra. Namely, four-point correlation function with one degenerate field 
\begin{equation}\label{SL2-4point}
  \xi(x,\bar{x})=
  \langle V_{-\frac{b}{2}}(x,\bar{x})V_{\alpha_1}(0)V_{\alpha_2}(1)V_{\alpha_3}(\infty)\rangle_L
\end{equation}
satisfies Fuchsian differential equations of the second order with three singular points in both variables $x$ and $\bar{x}$
\begin{equation}\label{DIFF2}
 \begin{aligned}
  &\left(\partial_x^2+\dots\right)\xi(x,\bar{x})=0,\\
  &\left(\partial_{\bar{x}}^2+\dots\right)\xi(x,\bar{x})=0.
 \end{aligned}
\end{equation}
Solutions of \eqref{DIFF2} are in general multi-valued function on a plane with three marked points $0$, $1$ and $\infty$ and one should find appropriate bilinear combination of solutions $\xi_i(x)$ and $\bar{\xi}_j(\bar{x})$
\begin{equation}
   \xi(x,\bar{x})=\sum_{ij}M_{ij}\xi_i(x)\bar{\xi}_j(\bar{x}),
\end{equation}
which is single-valued function. This condition fixes constants $M_{ij}$ up to normalization and in fact gives simple functional equation on the three-point correlation function (see for example ref \cite{Teschner:1995yf}). An expression for correlation function \eqref{SL2-4point} is given due to eqs \eqref{m-point} and \eqref{K1} by the two-dimensional Coulomb integral. 

A naive idea to generalize these arguments fails because corresponding correlation function in $\mathfrak{sl}(3)$ Toda theory
\begin{equation}\label{4-point-SL3}
  \langle V_{-b\omega_1}(x,\bar{x})V_{\alpha_1}(0)V_{\alpha_2}(1)V_{\alpha_3}(\infty)\rangle_T
\end{equation}
does no longer satisfy Fuchsian differential equation of the third order \cite{Fateev:2005gs,Bowcock:1993wq} and there is not known integral representation for this function in general case. Here we consider four-point function \eqref{4-point-SL3} with one semi-degenerate field $V_{\alpha_3}$, i.e. with $\alpha_3=\varkappa\omega_2-mb\omega_1$ and show that this function finally can be represented by the $4m+4$ dimensional integral. The case $m=0$ was considered in the first part of this paper \cite{Fateev:2007ab} and it was shown that correlation function  
\begin{equation}\label{4-point-SL3-m=0}
  \langle V_{-b\omega_1}(x,\bar{x})V_{\alpha_1}(0)V_{\alpha_2}(1)V_{\varkappa\omega_2}(\infty)\rangle_T
\end{equation}
can be represented by $4$ dimensional integral which is a solution of differential equation of the third order and can be expressed in terms of hypergeometric function $_{3}F_{2}(x)$. 

For convenience we define function $\Psi_{\alpha_1\alpha_2\alpha_3}(x,\bar{x})$ which is related with four-point correlation function with one completely degenerate field as
\begin{multline}\label{4point}
   \Psi_{\alpha_1\alpha_2\alpha_3}(x)=\bigl[\pi\mu\gamma(b^2)b^{2-2b^2}\bigr]^
   {\frac{(\alpha-b\omega_1-2Q,\rho)}{b}}\times\\\times 
   \frac{\langle V_{-b\omega_1}(x)V_{\alpha_1}(0)V_{\alpha_2}(1)V_{\alpha_3}(\infty)\rangle_T}
    {\prod_{e>0}\Upsilon\bigl((Q-\alpha_1,e)\bigr)\Upsilon\bigl((Q-\alpha_2,e)\bigr)
    \Upsilon\bigl((Q-\alpha_3,e)\bigr)}.
\end{multline}
This function  is again symmetric, Weyl invariant and $\mu$ independent function of variables $\alpha_1$, $\alpha_2$ and $\alpha_3$. Using integral relations proved in appendix one can show that function $\Psi_{\alpha_1\alpha_2\alpha_3}(x)$ satisfies integral relation reminding relations \eqref{Transformation2}
\begin{subequations}\label{Transformation3}
\begin{equation}
   \Psi_{\alpha_1\alpha_2\alpha_3}(x,\bar{x})=\int 
   \Psi_{\tilde{\alpha_1}\tilde{\alpha_2}\tilde{\alpha_3}}(y,\bar{y})\mathfrak{G}_{ijk}(x|y)\,d^2y,
\end{equation}
where 
\begin{multline}
  \mathfrak{G}_{ijk}(x|y)=\frac{1}{\gamma(b\varsigma_{ijk}-\frac{b^2}{3})}\\
       \frac{|x|^{2+2b^2+2b(\alpha_1-Q,h_i)}|x-1|^{2+2b^2+2b(\alpha_2-Q,h_j)}}
       {|y|^{2+2b^2+2b(\tilde{\alpha}_1-Q,h_i)}|y-1|^{2+2b^2+2b(\tilde{\alpha}_2-Q,h_j)}}
       |x-y|^{-2-2b\varsigma_{ijk}+\frac{2b^2}{3}},
\end{multline}
with
\begin{equation}
    \tilde{\alpha_1}=\alpha_1-\left(\varsigma_{ijk}+\frac{b}{3}\right)h_i,\qquad
    \tilde{\alpha_2}=\alpha_2-\left(\varsigma_{ijk}+\frac{b}{3}\right)h_j,\qquad
    \tilde{\alpha_3}=\alpha_3-\left(\varsigma_{ijk}+\frac{b}{3}\right)h_k
\end{equation}
\end{subequations}
with $\varsigma_{ijk}$ given by \eqref{sigma_{ijk}}. These relations give us again very useful tool to simplify four-point function in many cases. In particular, we can consider correlation function
\begin{equation}\label{Psi-xi112}
  \Psi_{\alpha_1,\alpha_2,\varkappa\omega_2-mb\omega_1+\varepsilon\omega_1}(x,\bar{x})=\int 
   \Psi_{\tilde{\alpha_1}\tilde{\alpha_2}\tilde{\alpha_3}}(y,\bar{y})\mathfrak{G}_{112}(x|y)\,d^2y 
\end{equation}
with
\begin{equation}
 \begin{aligned}
   &\tilde{\alpha}_1=\alpha_1-\left((\alpha_1-Q,h_1)+(\alpha_2-Q,h_1)+
    \frac{(\varkappa+(m+1)b-\varepsilon)}{3}\right)h_1,\\
   &\tilde{\alpha}_2=\alpha_2-\left((\alpha_1-Q,h_1)+(\alpha_2-Q,h_1)+
    \frac{(\varkappa+(m+1)b-\varepsilon)}{3}\right)h_1,\\
   &\tilde{\alpha}_3=\varkappa\omega_2-mb\omega_1+\varepsilon\omega_1
    -\left((\alpha_1-Q,h_1)+(\alpha_2-Q,h_1)+\frac{(\varkappa+(m+1)b-\varepsilon)}{3}\right)h_2.
 \end{aligned}
\end{equation}
In the limit $\varepsilon\rightarrow0$ we obtain
\begin{equation}
      (-b\omega_1+\tilde{\alpha_1}+\tilde{\alpha_2}+\tilde{\alpha_3}-2Q,\omega_1)=-(m+1)b,
\end{equation}
so we meet exactly the situation which is described by eq \eqref{Toda-Liouville-GL-relation}, i.e correlation function has a pole with residue given in terms of integral with Liouville correlation function. Namely, function $\Psi_{\tilde{\alpha_1}\tilde{\alpha_2}\tilde{\alpha_3}}(y,\bar{y})$ which enters in the r.h.s. of  \eqref{Psi-xi112} exhibits a pole when $\varepsilon\rightarrow0$ with residue\footnote{Field depending on the point $y$ does not appear in Liouville correlation function due to condition $(\omega_1,e_2)=0$.}
\begin{multline}
    \text{Res}\Bigl|_{\varepsilon=0}
    \Psi_{\tilde{\alpha_1}\tilde{\alpha_2}\tilde{\alpha_3}}(y,\bar{y})=
    \frac{\bigl[\pi\mu\gamma(b^2)b^{2-2b^2}\bigr]^{\frac{(\alpha-b\omega_1-2Q,\rho)}{b}}}
    {\prod_{e>0}\Upsilon\bigl((Q-\tilde{\alpha}_1,e)\bigr)\Upsilon\bigl((Q-\tilde{\alpha}_2,e)\bigr)
    \Upsilon\bigl((Q-\tilde{\alpha}_3,e)\bigr)}\times\\
    (-\pi\mu)^{m+1}\,|y|^{2b(\alpha_1,\omega_1)}\,|y-1|^{2b(\alpha_2,\omega_1)}
    \hspace*{-2.2pt}\int\prod_{i=1}^{m+1}|t_i|^{-3b(\alpha_1,\omega_1)}|t_i-1|^{-3b(\alpha_1,\omega_1)}
    |t_i-y|^{2b^2}\prod_{i<j}|t_i-t_j|^{-3b^2}\\\times
    \langle V_{-\frac{b}{2}}(t_1)\dots V_{-\frac{b}{2}}(t_{m+1})V_{\frac{(\tilde{\alpha}_1,e_2)}{2}}(0)
    V_{\frac{(\tilde{\alpha}_2,e_2)}{2}}(1)V_{\frac{(\tilde{\alpha}_3,e_2)}{2}}(\infty)\rangle_L\,
    d\mu_{m+1}(t).
\end{multline}
From the other side function $\Psi_{\alpha_1,\alpha_2,\varkappa\omega_2-mb\omega_1}(x,\bar{x})$ also has a simple pole due to the function $\Upsilon^{-1}((Q-\alpha_3,e_1))$ in its definition. Using integral representation for Liouville correlation function \eqref{m-point} we obtain for four-point correlation function in TFT with one semi-degenerate field the following expression
\begin{multline}\label{4point-final}
   \langle V_{-b\omega_1}(x,\bar{x})V_{\alpha_1}(0)V_{\alpha_2}(1)
   V_{\varkappa\omega_2-mb\omega_1}(\infty)\rangle_T=\Xi_{m+1}(\alpha_1,\alpha_2|\varkappa)
   \pi\frac{\gamma(1+\delta)\gamma((m+1)g)}{\gamma(\delta+(m+1)g)}\times\\\times
   |x|^{2b(\alpha_1,\omega_1)}\,|x-1|^{2b(\alpha_2,\omega_1)}\,
   \int |y-x|^{-2-2\delta}\boldsymbol{\mathfrak{S}}_{m+1}
   \left(\genfrac{}{}{0pt}{1}{A\;\;B}{A'\:B'}\,\boldsymbol{\Delta}\bigl|y\right)\,d^2y,
\end{multline}
where function $\boldsymbol{\mathfrak{S}}_{m+1}\left(\genfrac{}{}{0pt}{1}{A\;\;B}{A'\:B'}\,\boldsymbol{\Delta}\bigl|x\right)$ is given by eq \eqref{5-parametric-integral-x}, the factor $\Xi_{m+1}(\alpha_1,\alpha_2|\varkappa)$ is defined by eq \eqref{Xi-m} and
\begin{equation}
   A=\varrho_{32}^{m+1},\;\;
   B=\varrho_{23}^{m+1},\;\;
   A'=\varrho_{21}^{m+1},\;\;
   B'=\varrho_{12}^{m+1},\;\;
   \boldsymbol{\Delta}=1+\varrho_{22}^{m+1},
\end{equation}
here parameters $\varrho_{ij}^{m+1}$ are given by eq \eqref{varrho_{ij}^m} and 
\begin{equation}
   \delta=b(\alpha_1-Q,\omega_1)+b(\alpha_2-Q,\omega_1)+\frac{b(\varkappa+(m+1)b)}{3}.
\end{equation}
Applying relation \eqref{5parametric-reccurent} one can perform integration over variable $y$ and reduce four-point correlation function \eqref{4point-final} to $4(m+1)$ dimensional integral. 
%%%%%%%%%%%%%%%%%%%%%%%%%%%%%%%%%%%%%%%%%%%%%%%%%%%%%%%%%%%%%%%%%%%%%%%%%%%%%%%%%%%%%%%%%%%%%%%%%%%%%%%%%%%%%%%%%%%%%%%%%%%%%%%%%%%%%%%%%%%%%%%%%%%%%%%%%%%%%%%%%%%%%%%%%%%%%%%%%%%%%%%%%%%%%%%%%%%%%%%%%%%%%%%%%%%%%%%%%%%%%%%%%%%%%%%%%%%%%%%%%%%%%%%%%%%%%%%%%%%%%%%%%%%%%%%%%%%%%%%%%%%%%%%%%%%%%%%%%%%%%%%%%%%%%%%%%%%%%%%%%
\section{Concluding remarks}\label{Conclusion}
As was mentioned in the section \ref{SL3-Lagrangian}  three-point function $C(\alpha_1,\alpha_2,\alpha_3)$ considered as a function of $\alpha=\alpha_1+\alpha_2+\alpha_3$ has poles in the points
\begin{equation}\label{general-poles-1}
  (\alpha-2Q,\omega_1)=-mb-m'b^{-1},\qquad
  (\alpha-2Q,\omega_2)=-nb-n'b^{-1},
\end{equation} 
where $m$, $n$, $m'$ and $n'$ are some non-negative integers. Due to Weyl symmetry it follows from \eqref{general-poles-1} that three-point function has also poles in the points
\begin{equation}\label{general-poles-2}
   Q_L+\varsigma_{ijk}=-mb-m'b^{-1},\qquad
   Q_L-\varsigma_{ijk}=-nb-n'b^{-1},
\end{equation}
where $\varsigma_{ijk}$ is given by eq \eqref{sigma_{ijk}} and $Q_L$ by \eqref{QL}. We define function
\begin{equation}
   \boldsymbol{\mathfrak{Z}}(x)=\boldsymbol{G}(Q_L-x)\boldsymbol{G}(Q_L+x),
\end{equation}
where we introduce a self-dual entire function $\mathbf{G}(x)$ which contains only zeroes at the points $x=-nb-m/b$, $m,n=0,1,2,\ldots$ and enjoys the following shift relations
\begin{equation}
  \begin{aligned}
   &{\mathbf{G}}(x+b)=\frac{b^{1/2-bx}}{\sqrt{2\pi}}\Gamma(bx){\mathbf{G}}(x),\\
   &{\mathbf{G}}(x+1/b)=\frac{b^{x/b-1/2}}{\sqrt{2\pi}}\Gamma(x/b){\mathbf{G}}(x).
  \end{aligned}
\end{equation}
Evidently that function $\prod_{ijk}\boldsymbol{\mathfrak{Z}}^{-1}(\varsigma_{ijk})$ contains information about all poles \eqref{general-poles-2}. So, it is natural to consider the function
\begin{equation}
     \mathfrak{F}(\alpha_1,\alpha_2,\alpha_3)=
     C(\alpha_1,\alpha_2,\alpha_3)\prod_{ijk}\boldsymbol{\mathfrak{Z}}(\varsigma_{ijk})
\end{equation}
which is entire function\footnote{This statement has been checked in the case $\alpha_3=\varkappa\omega_2-mb\omega_1$, where this function is given by the integral representation.}. 
In Liouville field theory the similar product of three-point correlation function with eight $\mathbf{G}$ functions corresponding to Weyl transformed screening condition gives us the self-dual entire function which up to standard factor depending on $\mu$ is equal to $\Upsilon(2\alpha_1)\Upsilon(2\alpha_2)\Upsilon(2\alpha_3)$ where $\alpha_k$ are the parameters of the Liouville fields. In Toda theory function $\mathfrak{F}(\alpha_1,\alpha_2,\alpha_3)$ is very complicated entire function. The consideration of this function in the "light" semiclassical region and in the minisuperspace approximation (see ref \cite{Fateev:2007ab}) gives us the reasons to think that Weyl invariant function $\mathfrak{C}(\alpha_1,\alpha_2,\alpha_3)$ defined by eq \eqref{C-improved} has a simple poles only in the cases when up to Weyl transformation one of screening conditions \eqref{general-poles-2} is satisfied or if one  of the exponential fields is semidegenerate. The remarkable property \eqref{Transformation2} of the function $\mathfrak{C}(\alpha_1,\alpha_2,\alpha_3)$  relates the residues in the poles appearing from the screening condition with those coming from semidegenerate field. This property permits us to derive explicit integral representation for the three-point correlation function with one semidegenerate field. To go further in calculation of three-point function in TFT we need more information about analytical structure of entire function  $\mathfrak{F}(\alpha_1,\alpha_2,\alpha_3)$. We suppose to return  to the analysis of this function beyond semiclassical and minisuperspace approximations in near future.  
%%%%%%%%%%%%%%%%%%%%%%%%%%%%%%%%%%%%%%%%%%%%%%%%%%%%%%%%%%%%%%%%%%%%%%%%%%%%%%%%%%%%%%%%%%%%%%%%%%%%%%%%%%%%%%%%%%%%%%%%%%%%%%%%%%%%%%%%%%%%%%%%%%%%%%%%%%%%%%%%%%%%%%%%%%%%%%%%%%%%%%%%%%%%%%%%%%%%%%%%%%%%%%%%%%%%%%%%%%%%%%%%%%%%%%%%%%%%%%%%%%%%%%%%%%%%%%%%%%%%%%%%%%%%%%%%%%%%%%%%%%%%%%%%%%%%%%%%%%%%%%%%%%%%%%%%%%%%%%%%%
\acknowledgments
%%%%%%%%%%%%%%%%%%%%%%%%%%%%%%%%%%%%%%%%%%%%%%%%%%%%%%%%%%%%%%%%%%%%%%%%%%%%%%%%%%%%%%%%%%%%%%%%%%%%%%%%%%%%%%%%%%%%%%%%%%%%%%%%%%%%%%%%%%%%%%%%%%%%%%%%%%%%%%%%%%%%%%%%%%%%%%%%%%%%%%%%%%%%%%%%%%%%%%%%%%%%%%%%%%%%%%%%%%%%%%%%%%%%%%%%%%%%%%%%%%%%%%%%%%%%%%%%%%%%%%%%%%%%%%%%%%%%%%%%%%%%%%%%%%%%%%%%%%%%%%%%%%%%%%%%%%%%%%%% 
This work was supported, in part, by RBRF-CNRS grant PICS-09-02-91064. Work of A.~L. was supported  by DOE grant DE-FG02-96ER40949, by  RBRF grant 07-02-00799-a, by Russian Ministry of Science and Technology under the Scientific Schools grant 2044.2003.2 and by RAS program "Elementary particles and the fundamental nuclear physics". An important part of this paper has been made during the visits of A.~L. at the Laboratoire de Physique Th\'eorique et Astroparticules Universit\'e Montpellier~II within ENS-LANDAU program. V.~F. is very grateful to E.~Onofri for the numerical tests of the statements conjectured in the first part of this paper \cite{Fateev:2007ab}.
%%%%%%%%%%%%%%%%%%%%%%%%%%%%%%%%%%%%%%%%%%%%%%%%%%%%%%%%%%%%%%%%%%%%%%%%%%%%%%%%%%%%%%%%%%%%%%%%%%%%%%%%%%%%%%%%%%%%%%%%%%%%%%%%%%%%%%%%%%%%%%%%%%%%%%%%%%%%%%%%%%%%%%%%%%%%%%%%%%%%%%%%%%%%%%%%%%%%%%%%%%%%%%%%%%%%%%%%%%%%%%%%%%%%%%%%%%%%%%%%%%%%%%%%%%%%%%%%%%%%%%%%%%%%%%%%%%%%%%%%%%%%%%%%%%%%%%%%%%%%%%%%%%%%%%%%%%%%%%%%%
\appendix
%%%%%%%%%%%%%%%%%%%%%%%%%%%%%%%%%%%%%%%%%%%%%%%%%%%%%%%%%%%%%%%%%%%%%%%%%%%%%%%%%%%%%%%%%%%%%%%%%%%%%%%%%%%%%%%%%%%%%%%%%%%%%%%%%%%%%%%%%%%%%%%%%%%%%%%%%%%%%%%%%%%%%%%%%%%%%%%%%%%%%%%%%%%%%%%%%%%%%%%%%%%%%%%%%%%%%%
\section{Proof of the relations \protect\eqref{Transformation2} and \protect\eqref{Transformation3}.}\label{App-Integrals}
Both relations will be proved by using the same method. Namely, we assume for both correlation functions that screening condition is satisfied, i.e. 
\begin{equation}
   \sum_k\alpha_k+mbe_1+nbe_2=2Q. 
\end{equation}
In this case correlation function has a double pole with residue given by the $\mathfrak{sl}(3)$ Coulomb integral of dimension $2m+2n$. Using different identities we transform this integral to another $\mathfrak{sl}(3)$ integral and consider new one again as a residue of some correlation function. After that we assume that proved identity holds  not only for the residues but also for the entire correlation functions\footnote{Genrally this identity contains also some multiples of gamma functions, but hopfully these multiples always can be represented as a fraction of $\Upsilon$-functions.}. Of course, this is not a rigorous proof from the mathematical point of view and all statements proved in this manner have to be checked by another methods. All possible checks that we were able to make support the statements that are done in this appendix. 
%%%%%%%%%%%%%%%%%%%%%%%%%%%%%%%%%%%%%%%%%%%%%%%%%%%%%%%%%%%%%%%%%%%%%%%%%%%%%%%%%%%%%%%%%%%%%%%%%%%%%%%%%%%
\subsection*{Proof of the relations \protect\eqref{Transformation2}}
\begin{figure}
\psfrag{t}{$t$}\psfrag{s}{$s$}\psfrag{w}{$w$}\psfrag{m}{$m$}\psfrag{n}{$n$}\psfrag{k}{\hspace{10pt}$k$}
\psfrag{mk}{\hspace{-10pt}$m-k$}\psfrag{R}{\eqref{Nogi}}
	\centering
	\includegraphics[width=.9\textwidth]{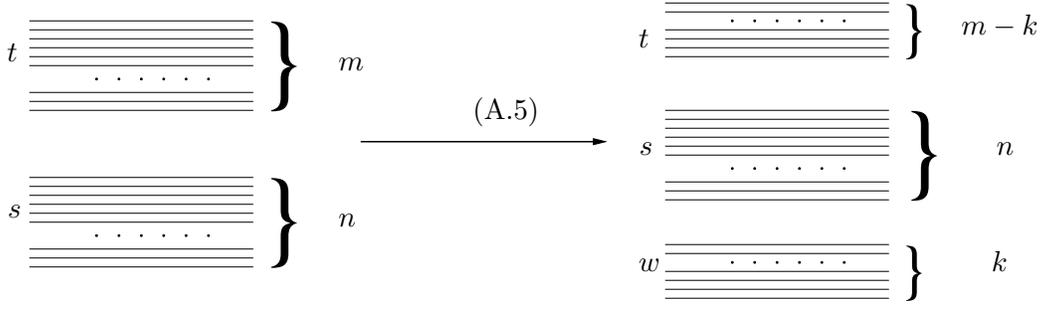}
	\caption{This picture represents integral property \protect\eqref{Nogi}. 
         In the l.h.s. we have a stack which consists of $m$ variables $t$ 
         and stack of $n$ variables $s$ with Coulomb interaction between 
         them associated with Cartan matrix of the Lie algebra $\mathfrak{sl}(3)$. In the r.h.s. we have 
         three stacks of variables: $m-k$ variables $t$, $n$ 
         variables $s$ and $k$ variables $w$ with Coulomb interaction associated 
         with the Lie algebra $\mathfrak{sl}(4)$.}
	\label{fig:nogi}
\end{figure}

Basic Coulomb integral for the  three-point correlation function \eqref{3point} due to \eqref{GoulLieFormula} is
\begin{equation}\label{Basic-3point-integral}
  \mathfrak{J}_{mn}(A_1,A_2,B_1,B_2|g)=
   \int \mathcal{I}_{mn}\left(\genfrac{}{}{0pt}{1}{A_1,\;\;B_1}{A_2,\;\;B_2}
\Bigl|\genfrac{}{}{0pt}{1}{t}{s}\right)d\mu_m(t)d\mu_n(s),
\end{equation}
where $A_k=-b(\alpha_1,e_k),\quad B_k=-b(\alpha_2,e_k),\quad g=-b^2$ and
\begin{equation}
   \mathcal{D}_m(t)=\prod_{i<j=1}^m|t_i-t_j|^2\qquad\text{and}\qquad
   d\mu_m(t)=\frac{1}{\pi^mm!}\prod_{k=1}^md^2t_k, 
\end{equation} 
and integrand is given by 
\begin{equation}\label{Imn}
   \mathcal{I}_{mn}\left(\genfrac{}{}{0pt}{1}{A_1,\;\;B_1}{A_2,\;\;B_2}
   \Bigl|\genfrac{}{}{0pt}{1}{t}{s}\right)=
   \prod_{i=1}^m|t_i|^{2A_1}|t_i-1|^{2B_1}\mathcal{D}_m^{2g}(t)
   \prod_{j=1}^n|s_j|^{2A_2}|s_j-1|^{2B_2}\mathcal{D}_n^{2g}(s)
   \prod_{ij}|t_i-s_j|^{-2g},
\end{equation}
For the future purposes  we prove basic property of this integral ("cross legs property"):
\begin{multline}\label{Nogi}
      \int \mathcal{I}_{m,\,n}\left(\genfrac{}{}{0pt}{1}{A_1,\;\;B_1}{A_2,\;\;B_2}
      \Bigl|\genfrac{}{}{0pt}{1}{t}{s}\right)d\mu_m(t)d\mu_n(s)=\Lambda_k(A_j,B_j)\times\\\times
      \int \prod_{j=1}^k|w_j|^{-4-2A_{12}-2(k-2)g}|w_j-1|^{-4-2B_{12}-2(k-2)g}\prod_{ij}|s_i-w_j|^{-2g}
      \mathcal{D}_{k}^{2g}(w)\cdot\\\cdot
      \mathcal{I}_{m-k,\,n}\left(\genfrac{}{}{0pt}{1}{A_1+kg,\;\;B_1+kg}{A_2,\;\;B_2}
      \Bigl|\genfrac{}{}{0pt}{1}{t}{s}\right)d\mu_{m-k}(t)\,d\mu_n(s)\,d\mu_k(w),
\end{multline}
where $A_{12}=A_1+A_2$, $B_{12}=B_1+B_2$ and
\begin{multline}
  \Lambda_k(A_j,B_j)=
   \prod_{j=0}^{k-1}\Biggl[\frac{\gamma((m-j)g)}{\gamma((j+1)g)}\times\\\times
   \frac{\gamma(1+A_1+jg)\gamma(1+B_1+jg)\gamma(2+A_{12}+(j-1)g)\gamma(2+B_{12}+(j-1)g)}
   {\gamma(2+A_1+B_1+(m-n-1+j)g)\gamma(3+A_{12}+B_{12}+(n-2+j)g)}\Biggr].
\end{multline}
This property allows to "move" part of the variables $t$ in the integral  through variables $s$ (see figure \ref{fig:nogi}).
To prove this property we will use integral relation  which was proposed in the paper \cite{Baseilhac:1998eq}
\begin{multline}\label{Bazeyaka}
   \int\mathcal{D}_n(y)\prod_{i=1}^n\prod_{j=1}^{n+m+1}|y_i-t_j|^{2p_j}
   \,d\mu_n(y)=\\=
   \frac{\prod_j\gamma(1+p_j)}{\gamma(1+n+\sum_jp_j)}
   \prod_{i<j}|t_i-t_j|^{2+2p_i+2p_j}%\times\\\times
   \int\mathcal{D}_m(u)\prod_{i=1}^m\prod_{j=1}^{n+m+1}|u_i-t_j|^{-2-2p_j}
   \,d\mu_m(u).
\end{multline}
The idea how to use this relation is rather simple and was explored widely in the paper \cite{Fateev:2007qn}. Namely, we start with the integral \eqref{Imn}  and represent for example $\mathcal{D}_m^{2g}(t)$ as
\begin{equation}\label{Start_Bazeyaka}
  \mathcal{D}^{2g}_m(t)=\mathcal{D}_m(t)
  \frac{\gamma(mg)}{\gamma^m(g)}
  \int\mathcal{D}_{m-1}(y)\prod_{i=1}^{m-1}\prod_{j=1}^{m}|y_i-t_j|^{-2+2g}d\mu_{m-1}(y).
\end{equation}
For simplicity in this appendix we will neglect further all numerical factors having in mind that all equations are valid up to these factors, which will be reconstructed in final expression.
After applying relation \eqref{Start_Bazeyaka} integral over variables $t$ is ready to be done again using relation \eqref{Bazeyaka}
\begin{multline}\label{Nogi-2step}
    \int\prod_{j=1}^m|t_j|^{2A_1}|t_j-1|^{2B_1}
    \prod_{i=1}^{m-1}|t_j-y_i|^{-2+2g}\prod_{k=1}^{n}|t_j-s_k|^{-2g}\mathcal{D}_{m}(t)\,d\mu_{m}(t)
    =\\=
    \prod_{i=1}^{m-1}|y_i|^{2(A_1+g)}\,|y_i-1|^{2(B_1+g)}\,\mathcal{D}_{m-1}^{2g-1}(y)\,
    \prod_{k=1}^{n}|s_k|^{2+2A_1-2g}\,|s_k-1|^{2+2B_1-2g}\,\mathcal{D}_n^{1-2g}(s)\times\\\times
    \int\prod_{j=1}^n|\nu_j|^{-2-2A_1}\,|\nu_j-1|^{-2-2B_1}
    \prod_{i=1}^{m-1}|\nu_j-y_i|^{-2g}\prod_{k=1}^n|\nu_j-s_k|^{-2+2g}
    \mathcal{D}_n(\nu)\,d\mu_n(\nu).
\end{multline}
We see that due to the factor $\mathcal{D}_n^{1-2g}(s)$ in \eqref{Nogi-2step} which combines with the factor $\mathcal{D}_n^{2g}(s)$ in \eqref{Imn}  we can perform now integration over variables $s$ using again relation \eqref{Bazeyaka}
\begin{multline}
     \int\prod_{j=1}^n|s_j|^{2+2A_{12}-2g}|s_j-1|^{2+2B_{12}-2g}
     \prod_{i=1}^n|s_j-\nu_i|^{-2+2g}
     \mathcal{D}_n(s)\,d\mu_n(s)=\mathcal{D}_{n}^{2g-1}(\nu)\times\\\times
     \prod_{j=1}^n|\nu_j|^{2+2A_{12}}|\nu_j-1|^{2+2B_{12}}
     %\times\\\times
     \int|w|^{-4-2A_{12}+2g}\,|w-1|^{-4-2B_{12}+2g}
     \prod_{i=1}^n|w-\nu_i|^{-2g}\,d\mu_1(w).
\end{multline}
Collecting all factors together finally we prove relation \eqref{Nogi} for the case $k=1$. Repeating similar steps we can prove \eqref{Nogi} for $k=2$ and so on. For us will be important identity \eqref{Nogi} with $k=m$. If we suppose that this identity also holds "away" from the screening contition it gives functional relation for function $\mathfrak{C}(\alpha_1,\alpha_2,\alpha_3)$ defined by eq \eqref{C-improved}
\begin{equation}
     \mathfrak{C}(\alpha_1,\alpha_2,\alpha_3)=
     \mathfrak{C}(\tilde{\alpha}_1,\tilde{\alpha}_2,\tilde{\alpha}_3),
\end{equation}
where
\begin{equation}\label{App-tilde-alpha}
    \tilde{\alpha}_k=\alpha_k^{*}+(\alpha-3\alpha_k,\omega_1)\omega_2
\end{equation}
with parameter $\alpha_k^{*}$ defined by relation $(\alpha_k^{*},e_k)=(\alpha_k,e_{3-k})$.
Now if one applies Weyl reflection $\boldsymbol{\sigma_1}\boldsymbol{\sigma_2}$ to each $\tilde{\alpha_k}$ one gets
\begin{equation}
    \boldsymbol{\sigma_1}\boldsymbol{\sigma_2}(\tilde{\alpha}_k)=\alpha_k-\varsigma_{111}h_1,
\end{equation}
where $\varsigma_{111}$ is given by \eqref{sigma_{ijk}}. Other relation for arbitrary $i$, $j$ and $k$ can be obtained from this one by Weyl reflections. 

Here we also give without a proof another "cross legs property"  of the integral \eqref{Basic-3point-integral} which also gives very useful tool for studying  properties of correlation functions in TFT  
\begin{multline}\label{Nogi-shifted}
      \int \mathcal{I}_{m,\,n}\left(\genfrac{}{}{0pt}{1}{A_1,\;\;B_1}{A_2,\;\;B_2}
      \Bigl|\genfrac{}{}{0pt}{1}{t}{s}\right)d\mu_m(t)d\mu_n(s)=\tilde{\Lambda}_k(A_j,B_j)\times\\\times
      \int \prod_{j=1}^k|w_j|^{-4-2A_{12}-2(m-2)g}|w_j-1|^{-4-2B_{12}-2(k-2)g}\prod_{ij}|s_i-w_j|^{-2g}
      \mathcal{D}_{k}^{2g}(w)\cdot\\\cdot
      \mathcal{I}_{m-k,\,n}\left(\genfrac{}{}{0pt}{1}{A_1,\;\;B_1+kg}{A_2,\;\;B_2}
      \Bigl|\genfrac{}{}{0pt}{1}{t}{s}\right)d\mu_{m-k}(t)\,d\mu_n(s)\,d\mu_k(w),
\end{multline}
where $\tilde{\Lambda}_k(A_1,A_2,B_1,B_2)\overset{\text{def}}{=}\Lambda_k(A_1+(m-k)g,A_2,B_1,B_2)$. We note that for $k=m$ transformation coincide with \eqref{Nogi}. We see also that \eqref{Nogi-shifted} is not symmetric with respect of exchange of points  $0$ and $1$. Of course there is another relation similar to \eqref{Nogi-shifted} but with substitution $0\rightarrow1$.
%%%%%%%%%%%%%%%%%%%%%%%%%%%%%%%%%%%%%%%%%%%%%%%%%%%%%%%%%%%%%%%%%%%%%%%%%%%%%%%%%%%%%%%%%%%%%%%%%%%%%%%%%%%
\subsection*{Proof of the relations \protect\eqref{Transformation3}}
Basic integral for the four-point correlation function with one degenerate field 
$$\langle V_{-b\omega_1}(x,\bar{x})V_{\alpha_1}(0)V_{\alpha_2}(1)V_{\alpha_3}(\infty)\rangle$$ is 
\begin{equation}
   \int \prod_{j=1}^m|t_j-x|^{-2g}\,
   \mathcal{I}_{mn}\left(\genfrac{}{}{0pt}{1}{A_1,\;\;B_1}{A_2,\;\;B_2}
   \Bigl|\genfrac{}{}{0pt}{1}{t}{s}\right)d\mu_m(t)d\mu_n(s).
\end{equation}
One can prove "cross legs" property for this integral:
\begin{multline}\label{Nogi-4point}
      \int\prod_{j=1}^m|t_j-x|^{-2g}\, 
      \mathcal{I}_{m,\,n}\left(\genfrac{}{}{0pt}{1}{A_1,\;\;B_1}{A_2,\;\;B_2}
      \Bigl|\genfrac{}{}{0pt}{1}{t}{s}\right)d\mu_m(t)d\mu_n(s)=\Lambda_{m-1}(A_j,B_j)\times\\\times
      \int \prod_{j=1}^{m-1}|w_j|^{-4-2A_{12}-2(m-3)g}|w_j-1|^{-4-2B_{12}-2(m-3)g}\prod_{ij}|s_i-w_j|^{-2g}
      \mathcal{D}_{m-1}^{2g}(w)\cdot\\\cdot
      |t-x|^{-2mg}
      \mathcal{I}_{1,\,n}\left(\genfrac{}{}{0pt}{1}{A_1+(m-1)g,\;\;B_1+(m-1)g}{A_2,\;\;B_2}
      \Bigl|\genfrac{}{}{0pt}{1}{t}{s}\right)d\mu_{1}(t)\,d\mu_n(s)\,d\mu_{m-1}(w),
\end{multline}
We see that in the case of four-point correlation function one can "move" only $(m-1)$ variables $t$ but not arbitrary number as it was in the case of three-point function. This identity can be proved using the same technique but the proof is a little bit more involved. To make the calculations more transparent we consider the case $m=2$. The general case follows the same steps but calculations are more involved. So, we consider integral
\begin{equation}\label{App-4point-m=2}
   \int \prod_{j=1}^2|t_j-x|^{-2g}\,
   \mathcal{I}_{2,\,n}\left(\genfrac{}{}{0pt}{1}{A_1,\;\;B_1}{A_2,\;\;B_2}
   \Bigl|\genfrac{}{}{0pt}{1}{t}{s}\right)d\mu_2(t)d\mu_n(s)
\end{equation}
As usual we start by representing
\begin{equation}
    \mathcal{D}_{2}^{2g}(t)=\mathcal{D}_{2}(t)\int\prod_{i=1,2}|\xi-t_i|^{-2+2g}\,d\mu_1(\xi),
\end{equation}
and after that we perform integration over variables $t$ with a result
\begin{multline}\label{int-step2}
   \int\prod_{j=1}^2|t_j|^{2A_1}\,|t_j-1|^{2B_1}\,|t_j-x|^{-2g}
   |t_j-\xi|^{-2+2g}\prod_{i=1}^n|t_j-s_i|^{-2g}\mathcal{D}_2(t)\,d\mu_2(t)=\mathcal{D}_{n}^{1-2g}(s)
   \times\\\prod_{i=1}^n|s|^{2+2A_1-2g}|s_i-1|^{2+2B_1-2g}|s_i-x|^{2-4g}
   |x|^{2+2A_1-2g}|x-1|^{2+2B_1-2g}|\xi|^{2A_1+2g}|\xi-1|^{2B_1+2g}%\times\\\times
   \\%\times
   \int\prod_{j=1}^{n+1}|\eta_j|^{-2-2A_1}\,|\eta_j|^{-2-2B_1}\,
   |\eta_j-x|^{-2+2g}\,|\eta_j-\xi|^{-2g}
   \prod_{i=1}^n|\eta_j-s_i|^{-2+2g}\mathcal{D}_{n+1}(\eta)\,d\mu_{n+1}(\eta).
\end{multline}
We see that again factor $\mathcal{D}_{n}^{1-2g}(s)$ in \eqref{int-step2} permits us to perform integration over variables $s$:
\begin{multline}
   \int\prod_{j=1}^n|s_j|^{2+2A_{12}-2g}\,|s_j-1|^{2+2B_{12}-2g}\,
   |s_j-x|^{2-4g}\prod_{i=1}^{n+1}|s_j-\eta_i|^{-2+2g}\mathcal{D}_n(s)\,d\mu_n(s)
   =\\=
   |x|^{6+2A_{12}-6g}\,|x-1|^{6+2B_{12}-6g}%\times\\\times
   \prod_{j=1}^{n+1}|\eta_j|^{2+2A_{12}}
   |\eta_j-1|^{2+2B_{12}}\,|\eta_j-x|^{2-2g}\,\mathcal{D}_{n+1}^{2g-1}(\eta)
   \times\\\times
   \int\prod_{i=1}^3|w_i|^{-4-2A_{12}+2g}|w_i-1|^{-4-2B_{12}+2g}\,|w_i-x|^{-4+4g}
   \prod_{j=1}^{n+1}|w_i-\eta_j|^{-2g}
   \mathcal{D}_{3}(w)\,d\mu_3(w).
\end{multline}
At this point we have integrated over all variables $t$ and $s$ of the original integral \eqref{App-4point-m=2}. So we proved that integral \eqref{App-4point-m=2} equals to
\begin{multline}
   |x|^{8+4A_1+2A_2-8g}\,|x-1|^{8+4B_1+2B_2-8g}
   \int\mathcal{I}_{1,\,n+1}\left(\genfrac{}{}{0pt}{1}{A_1+g,\;\;B_1+g}{A_2,\;\;B_2}
   \Bigl|\genfrac{}{}{0pt}{1}{\xi}{\eta}\right)\\\hspace*{-5pt}\prod_{j=1}^3\hspace*{-2pt}
   |w_j|^{-4-2A_{12}+2g}|w_j-1|^{-4-2B_{12}+2g}|w_j-x|^{-4+4g}\hspace*{-3pt}
   \prod_{i=1}^{n+1}|w_j-\eta_i|^{-2g}\mathcal{D}_{3}(w)d\mu_1(\xi)d\mu_n(\eta)d\mu_3(w)
\end{multline}
The second step is again to integrate over all new variables $\xi$, $\eta$ and $w$. First we integrate over variables $\xi$
\begin{multline}
   \int|\xi|^{2A_1+2g}\,|\xi-1|^{2B_1+2g}\prod_{j=1}^{n+1}|\xi-\eta_j|^{-2g}\,d\mu_{1}(\xi)=
   \prod_{j=1}^{n+1}|\eta_j|^{2+2A_1}\,|\eta_j-1|^{2+2B_1}\,\mathcal{D}_{n+1}^{1-2g}(\eta)
   \times\\\times
   \int
   \prod_{i=1}^{n+1}|\nu_i|^{-2-2A_1-2g}\,|\nu_i-1|^{-2-2B_1-2g}
   \prod_{j=1}^{n+1}|\nu_i-\eta_j|^{-2+2g}\mathcal{D}_{n+1}(\nu)\,d\mu_{n+1}(\nu),
\end{multline}
and after that we perform integration over variables $\eta$
\begin{multline}\label{int-step3}
     \int\prod_{j=1}^{n+1}|\eta_j|^{2+2A_{12}}\,|\eta_j-1|^{2+2B_{12}}
     \prod_{i=1}^{n+1}|\eta_j-\nu_i|^{-2+2g}
     \prod_{k=1}^3|\eta_j-w_k|^{-2g}\mathcal{D}_{n+1}(\eta)\,d\mu_{n+1}(\eta)
     =\\=\prod_{i=1}^{n+1}|\nu_i|^{2+2A_{12}+2g}\,|\nu_i-1|^{2+2B_{12}+2g}\,\mathcal{D}_{n+1}^{2g-1}(\nu)\,
     \prod_{k=1}^3|w_k|^{4+2A_{12}-2g}\,|w_k-1|^{4+2B_{12}-2g}\,\mathcal{D}_{3}^{1-2g}(w)\\\times
     \int\prod_{j=1}^4|\tau_j|^{-4-2A_{12}}\,|\tau_j-1|^{-4-2B_{12}}\,
     \prod_{i=1}^{n+1}|\tau_j-\nu_i|^{-2g}\prod_{k=1}^3|\tau_j-w_k|^{-2+2g}
     \mathcal{D}_4(\tau)\,d\mu_4(\tau).
\end{multline}
At this moment we meet first obstruction with the integral over variables $w$ which is not of the type we met before
\begin{equation}\label{App-Int-Phi13-Phi12x4}
    J(x,\tau_i|g)=\int\prod_{j=1}^3|w_j-x|^{-4+4g}\prod_{k=1}^4|w_j-\tau_i|^{-2+2g}\mathcal{D}_3^{2-2g}(w)
    d\mu_3(w),
\end{equation}
one can show that this integral equals to one dimensional integral\footnote{This integral identity reflects the fact that correlation function of degenerate fields in Liouville CFT can be represented by integrals with different number of screening charges. In eqs \eqref{App-Int-Phi13-Phi12x4} and \eqref{App-Int-Phi13-Phi12x4-Identity} the coupling constant is given by $b^2=g-1$.}
\begin{equation}\label{App-Int-Phi13-Phi12x4-Identity}
   J(x,\tau_i|g)=\frac{\gamma^3(g)(1-3g)^2}{\gamma(3g)(1-2g)^2}
    \prod_{j=1}^4|\tau_j-x|^{-4+6g}
    \int|\rho-x|^{4-8g}\prod_{j=1}^4|\rho-\tau_j|^{-2+2g}d\mu_1(\rho).
\end{equation}
Now we can perform integral over variables $\tau$:
\begin{multline}
    \int\prod_{j=1}^4|\tau_j|^{-4-2A_{12}}|\tau_j-1|^{-4-2B_{12}}\,|\tau_j-x|^{-4+6g}\,
    \prod_{i=1}^{n+1}|\tau_j-\nu_i|^{-2g}\,|\tau_j-\rho|^{-2+2g}\mathcal{D}_4(\tau)\,d\mu_4(\tau)
    =\\|x|^{-6-2A_{12}+6g}\,|x-1|^{-6-2B_{12}+6g}\,
     \prod_{i=1}^{n+1}|\nu_i|^{-2-2A_{12}-2g}\,|\nu_i-1|^{-2-2B_{12}-2g}
    \,|\nu_i-x|^{-2+4g}\mathcal{D}_{n+1}^{1-2g}(\nu)\times\\\times
   |\rho|^{-4-2A_{12}+2g}\,|\rho-1|^{-4-2B_{12}+2g}\,|\rho-x|^{-4+8g}\times\\\times
   \int\prod_{j=1}^n|\lambda_j|^{2+2A_{12}}\,|\lambda_j-1|^{2+2B_{12}}\,|\lambda_j-x|^{2-6g}
   \prod_{i=1}^{n+1}|\lambda_j-\nu_i|^{-2+2g}\,|\lambda_j-\rho|^{-2g}\mathcal{D}_n(\lambda)d\mu_n(\lambda),
\end{multline}
It follows from eqs \eqref{int-step3}, \eqref{App-Int-Phi13-Phi12x4} and \eqref{App-Int-Phi13-Phi12x4-Identity} that factor $\mathcal{D}_{n+1}(\nu)$ appears in the first power and we can  perform integral over variables $\nu$:
\begin{multline}
   \int\prod_{j=1}^{n+1}|\nu_j|^{-2-2A_1-2g}|\nu_j-1|^{-2-2B_1-2g}|\nu_j-x|^{-2+4g}
   \prod_{i=1}^n|\nu_j-\lambda_i|^{-2+2g}\mathcal{D}_{n+1}(\nu)d\mu_{n+1}(\lambda)=\\=
   |x|^{-2-2A_1+2g}\,|x-1|^{-2-2B_1+2g}
   \prod_{j=1}^n|\lambda_j|^{-2-2A_1}|\lambda_j-1|^{-2-2B_1}|\lambda_j-x|^{-2+6g}
   \mathcal{D}_{n}^{2g-1}(\lambda)\times\\\times
   \int|\zeta|^{2A_1+2g}|\zeta-1|^{2B_1+2g}|\zeta-x|^{-4g}\prod_{j=1}^n|\zeta-\lambda_j|^{-2g}
   d\mu_1(\zeta).
\end{multline}
Collecting now all missing factors we obtain \eqref{Nogi-4point} for $m=2$. In a similar way using identities like \eqref{App-Int-Phi13-Phi12x4-Identity} we can prove \eqref{Nogi-4point} for arbitrary $m$.
Relation \eqref{Nogi-4point} being analytically continued to the non-integer number of screenings gives functional equation for function $\Psi_{\alpha_1\alpha_2\alpha_3}(x,\bar{x})$ defined by eq \eqref{4point}
%%%%%%%%%%%%%%%%%%%%%%%%%%%%%%%%%%%%%%%%%%%%%%%%%%%%%%%%%%%%%%%%%%%%%%%%%%%%%%%%%%%%%%%%%%%%%%%%%%%%%%%%%%%%%%%%%%%%%%%%%%%%%%%%%%%%%%%%%%%%%%%%%%%%%%%%%%%%%%%%%%%%%%%%%%%%%%%%%%%%%%%%%%%%%%%%%%%%%%%%%%%%%%%%%%%%%%
\section{Properties of the kernel}\label{Kernel}
In this appendix we define function $K_m^{\Delta}(t|y)$ which enters in the integral representation for the Liouville correlation function \eqref{m-point}. This function was defined in ref \cite{Fateev:2007qn}. It is symmetric function of variables $t_k$ and $y_k$ which does not change under the permutation $t_k\leftrightarrow y_k$. The last two properties are not evident from the explicit form of the function $K_m^{\Delta}(t|y)$, but they can be proved. This function is given by the $m(m-1)$-dimensional Coulomb integral and can be derived from the recurrent equation\footnote{To make sense of this formula we set $K_0^{\Delta}=1$.}
\begin{multline}\label{Kernel-recurrent}
   K_m^{\boldsymbol{\Delta}}(t_1,..,t_m|y_1,..,y_m)=
   \frac{\gamma(mg)}{\gamma^m(g)}\,\mathcal{D}_m^{1-2g}(t)\;
   \prod_{k=1}^m|t_k-y_1|^{-2\boldsymbol{\Delta}}
   \times\\\times
   \int\mathcal{D}_{m-1}(\tau)\prod_{j=1}^{m-1}|\tau_j-y_1|^{2(\boldsymbol{\Delta}-g)}
   \prod_{k=1}^{m}|\tau_j-t_k|^{-2+2g}
   K_{m-1}^{\boldsymbol{\Delta}+g}(\tau_1,..,\tau_{m-1}|y_2,..,y_m)\,d\mu_{m-1}(\tau).
\end{multline}
The simplest example of the kernel is the case $m=1$:
\begin{equation}\label{K1}
    K_1(t|y)=|t-y|^{-2\boldsymbol{\Delta}}.
\end{equation}

We consider basic integral which appears in eq \eqref{C-m} and has a form 
\begin{multline}\label{5-parametric-integral}
 \mathfrak{I}_m\left(\genfrac{}{}{0pt}{1}{A\;\;B}{A'\:B'}\,\boldsymbol{\Delta}\right)=\\=
 \int \prod_{k=1}^m|t_k|^{2A}|t_k-1|^{2B}|s_k|^{2A'}|s_k-1|^{2B'}
 K_m^{\boldsymbol{\Delta}}(t|s)\mathcal{D}_m^{2g}(t)\mathcal{D}_m^{2g}(s)
 d\mu_m(t)\,d\mu_m(s).
\end{multline}
The simplest case $m=1$
\begin{equation}\label{5-parametric-integral-m=1}
 \mathfrak{I}_1\left(\genfrac{}{}{0pt}{1}{A\;\;B}{A'\:B'}\,\boldsymbol{\Delta}\right)=
 \int|t|^{2A}|t-1|^{2B}|s|^{2A'}|s-1|^{2B'}|t-s|^{-2\boldsymbol{\Delta}}
 d^2t\,d^2s.
\end{equation}
This integral can be calculated exactly and expressed in terms of hypergeometric function ${}_3F_2(1)$\footnote{This integral is particular case corresponding to $n=3$ of more general Coulomb integral associated with Lie algebra $\mathfrak{sl}(n)$ calculated in \cite{Fateev:2007ab}.}. Another important integral with function $K_m^{\Delta}(t_1,..,t_m|y_1,..,y_m)$ which appears in eq \eqref{4point-final} has a form
\begin{multline}\label{5-parametric-integral-x}
  \boldsymbol{\mathfrak{S}}_{m}\left(\genfrac{}{}{0pt}{1}{A\;\;B}{A'\:B'}\Bigl|
  \boldsymbol{\Delta},x\right)=\\
  \int \prod_{k=1}^m|t_k|^{2A}|t_k-1|^{2B}|s_k|^{2A'}|s_k-1|^{2B'}|t_k-x|^{-2g}
  K_m^{\boldsymbol{\Delta}}(t|s)\mathcal{D}_m^{2g}(t)\mathcal{D}_m^{2g}(s)
  d\mu_m(t)\,d\mu_m(s).
\end{multline}
Using factorization property (B.3) from ref \cite{Fateev:2007qn} and using relation \eqref{Nogi-4point} we can prove that it satisfies an important recurrent identity
\begin{multline}\label{5parametric-reccurent}
   \boldsymbol{\mathfrak{S}}_{m+1}\left(\genfrac{}{}{0pt}{1}{A\;\;B}{\,A'\:B'}\,
   \boldsymbol{\Delta}\Bigl|x\right)=\Theta_m\left(\genfrac{}{}{0pt}{1}{A\;\;B}{\,A'\:B'}\,
   \boldsymbol{\Delta}\right)\times\\\times
   \int |t|^{2(A+mg)}|t-1|^{2(B+mg)}\,|y|^{2(A'+mg)}|y-1|^{2(B'+mg)}\,
   |t-y|^{-2(\boldsymbol{\Delta}+mg)}\,|t-x|^{-2(m+1)g}\cdot\\\cdot
   \boldsymbol{\mathfrak{S}}_{m}\left(\genfrac{}{}{0pt}{1}
   {\,\,\tilde{A}'\:\tilde{B}'}{\tilde{A}\;\,\tilde{B}}\,
   \tilde{\boldsymbol{\Delta}}\Bigl|y\right)d^2t\,d^2y
\end{multline}
where 
\begin{equation}
\begin{gathered}
 \begin{aligned}
    &\tilde{A}=-1-A-(m-1)g,&\qquad &\tilde{B}=-1-B-(m-1)g,\\
    &\tilde{A}'=-1-A'-(m-1)g,&\qquad &\tilde{B}'=-1-B'-(m-1)g,
 \end{aligned}\\
  \tilde{\boldsymbol{\Delta}}=-1-\boldsymbol{\Delta}-(m-1)g 
\end{gathered}
\end{equation}
and
\begin{multline}
  \Theta_m\left(\genfrac{}{}{0pt}{1}{A\;\;B}{\,A'\:B'}\,\boldsymbol{\Delta}\right)=\\
   =\prod_{j=0}^{m-1}
  \frac{\gamma(1+A+jg)\gamma(1+B+jg)\gamma(1+A'+jg)\gamma(1+A'+jg)\gamma(1-\boldsymbol{\Delta}-jg)}
  {\gamma(2+A+B-\boldsymbol{\Delta}+jg)\gamma(2+A'+B'-\boldsymbol{\Delta}+(1-j)g)}\times\\\times
  \frac{\gamma(2+A+A'-\boldsymbol{\Delta}+jg)\gamma(2+B+B'-\boldsymbol{\Delta}+jg)}
  {\gamma(3+A+B+A'+B'-\boldsymbol{\Delta}+(j+1)g)}.
\end{multline}
We note, that we can apply relation \eqref{5parametric-reccurent} to function $\boldsymbol{\mathfrak{S}}_{m}\left(\genfrac{}{}{0pt}{1}{\,\,\tilde{A}'\:\tilde{B}'}{\tilde{A}\;\,\tilde{B}}\,\tilde{\boldsymbol{\Delta}}\Bigl|y\right)$ again and finally reduce integral \eqref{5-parametric-integral-x} to $4m$-dimensional integral. It is easy to see that due to eq \eqref{5parametric-reccurent} the dependence of the integral \eqref{5-parametric-integral-x} on variable $x$ is simple and it permits us to perform integration in \eqref{4point-final}.
 
Integral \eqref{5-parametric-integral} can be obtained from the integral \eqref{5-parametric-integral-x} by considering the limit at $x\rightarrow\infty$
\begin{equation}\label{limit}
   \lim_{x\rightarrow\infty}|x|^{2mg} 
   \boldsymbol{\mathfrak{S}}_{m}\left(\genfrac{}{}{0pt}{1}{A\;\;B}{\,A'\:B'}\,
   \boldsymbol{\Delta}\Bigl|x\right)=
   \mathfrak{I}_m\left(\genfrac{}{}{0pt}{1}{A\;\;B}{A'\:B'}\biggl|\boldsymbol{\Delta}\right).
\end{equation}
Taking the limit \eqref{limit} in both sides of \eqref{5parametric-reccurent} we obtain 
\begin{multline}\label{5parametric-reccurent-1}
   \mathfrak{I}_{m+1}\left(\genfrac{}{}{0pt}{1}{A\;\;B}{\,A'\:B'}\,
   \boldsymbol{\Delta}\right)=\Theta_m\left(\genfrac{}{}{0pt}{1}{A\;\;B}{\,A'\:B'}\,
   \boldsymbol{\Delta}\right)\times\\\times
   \int |t|^{2(A+mg)}|t-1|^{2(B+mg)}\,|y|^{2(A'+mg)}|y-1|^{2(B'+mg)}\,
   |t-y|^{-2(\boldsymbol{\Delta}+mg)}\cdot\\\cdot
   \boldsymbol{\mathfrak{S}}_{m}\left(\genfrac{}{}{0pt}{1}
   {\,\,\tilde{A}'\:\tilde{B}'}{\tilde{A}\;\,\tilde{B}}\,
   \tilde{\boldsymbol{\Delta}}\Bigl|y\right)d^2t\,d^2y.
\end{multline}
Now we can apply recurrent relation \eqref{5parametric-reccurent} to the r.h.s. of \eqref{5parametric-reccurent-1} and finally we reduce integral $\mathfrak{I}_m\left(\genfrac{}{}{0pt}{1}{A\;\;B}{A'\:B'}\biggl|\boldsymbol{\Delta}\right)$ defined by eq \eqref{5-parametric-integral} to $4m$-dimensional integral.
%%%%%%%%%%%%%%%%%%%%%%%%%%%%%%%%%%%%%%%%%%%%%%%%%%%%%%%%%%%%%%%%%%%%%%%%%%%%%%%%%%%%%%%%%%%%%%%%%%%%%%%%%%%%%%%%%%%%%%%%%%%%%%%%%%%%%%%%%%%%%%%%%%%%%%%%%%%%%%%%%%%%%%%%%%%%%%%%%%%%%%%%%%%%%%%%%%%%%%%%%%%%%%%%%%%%%%%%%%%%%%%%%%%%%%%%%%%%%%%%%%%%%%%%%%%%%%%%%%%%%%%%%%%%%%%%%%%%%%%%%%%%%%%%%%%%%%%%%%%%%%%%%%%%%%%%%%%%%%%%%
\bibliographystyle{JHEP} 
\bibliography{MyBib}
\end{document}